\documentclass[10pt]{article}
\usepackage{graphicx}
\usepackage{amsmath}
\usepackage{amssymb}
\usepackage{caption2}
\begin{document}
\begin{center}
{\large {\bf \sc{  Analysis of  the scalar nonet mesons  with QCD sum rules }}} \\[2mm]
    Zhi-Gang Wang\footnote{ E-mail,zgwang@aliyun.com.  }     \\
  Department of Physics, North China Electric Power University, Baoding 071003, P. R. China \\
\end{center}

\begin{abstract}
In this article, we assume that the nonet scalar mesons  below $1\,\rm{ GeV}$ are the two-quark-tetraquark  mixed states and study their  masses and pole residues  using the  QCD sum rules. In calculation, we take into account the vacuum condensates up to dimension 10 and the $\mathcal{O}(\alpha_s)$ corrections to the perturbative   terms in the operator product expansion.   We determine the mixing angles, which indicate  the two-quark components are much  larger than $50\%$, then obtain the masses and  pole residues of the nonet scalar mesons.
\end{abstract}



\section{Introduction}

There are many scalar mesons  below $2\,\rm{GeV}$, which
cannot be accommodated in one $\bar{q}q$ nonet, some are supposed to
be glueballs, molecular states and tetraquark states \cite{PDG,Close2002,ReviewJaffe,ReviewAmsler2,KlemptPRT}.
In the scenario of molecular states, the  scalar states below $1\,\rm{GeV}$ are taken as loosely bound mesonic molecular states \cite{Molecule}, or dynamical generated resonances \cite{Dynami}.
On the other hand, in the scenario of tetraquark states, if we suppose  the dynamics dominates  the scalar
 mesons  below and above $1\,\rm{GeV}$ are different,  there maybe exist  two scalar nonets below $1.7\,\rm{ GeV}$ \cite{Close2002,ReviewJaffe,ReviewAmsler2}.
  The strong attractions between the scalar diquarks  and  anti-diquarks in relative $S$-wave maybe  result in a
nonet tetraquark states  manifest below $1\,\rm{GeV}$, while the conventional $^3P_0$ quark-antiquark nonet mesons have masses about $(1.2-1.6)
\,\rm{GeV}$. The well established  $^3P_1$ and $^3P_2$ quark-antiquark nonets  lie in the same region.  In 2013, Weinberg explored the tetraquark states in the large-$N_c$ limit    and observed that the existence of light tetraquark states is consistent with large-$N_c$
QCD \cite{Weinberg}.
We usually take  the lowest scalar  nonet  mesons $\{f_0/\sigma(500),a_0(980),\kappa_0(800),f_0(980) \}$ to be the tetraquark
states,  and assign the higher  scalar nonet mesons
$\{f_0(1370),a_0(1450),K^*_0(1430),f_0(1500) \}$ to be the
conventional ${}^3P_0$ quark-antiquark  states \cite{Close2002,ReviewJaffe,ReviewAmsler2,TetraQuark}.

There maybe exist some mixing between the two scalar nonet mesons, for example, in the chiral theory \cite{Giacosa}.
In the  naive quark  model, for $f_0(980)= \bar{s}s$, the strong decay   $f_0(980)\to \pi\pi$ is Okubo-Zweig-Iizuka forbidden; for $a_0^0(980)={u\bar{u}-d\bar{d}\over\sqrt{2}}$, the radiative decay $\phi(1020)\to a_0^0(980)\gamma$ is both Okubo-Zweig-Iizuka forbidden and isospin violated. From the Review of Particle Physics, we can see that the process $f_0(980)\to \pi\pi$ dominates  the decays of the $f_0(980)$ and the branching fractions ${\rm Br}\left(\,\phi(1020)\to a_0^0(980)\gamma\,\right)=
(7.6 \pm 0.6 ) \times 10^{-5}$, ${\rm Br}\left(\,\phi(1020)\to f_0(980)\gamma\,\right)=
(3.22 \pm 0.19 ) \times 10^{-4}$ \cite{PDG}. The naive quark model cannot account for the experimental data even qualitatively, we have to introduce some tetraquark constituents, such as ${us\bar{u}\bar{s}+ds\bar{d}\bar{s}\over\sqrt{2}}$ and ${us\bar{u}\bar{s}-ds\bar{d}\bar{s}\over\sqrt{2}}$, if we do not want to turn on the instanton effects  \cite{Hooft}.

We can use  QCD sum rules to study the two-quark and tetraquark states.
QCD sum rules provides  a powerful theoretical tool  in
 studying the hadronic properties, and has been applied extensively
  to study the masses, decay constants, hadronic form-factors,  coupling constants, etc  \cite{SVZ79,PRT85}.
There have been several works on the light tetraquark states using the QCD sum rules
\cite{Nielsen2005,Wang-nonet,Wang-scalar,WangNPA,Lee2006,Chen-Zhu,tetra-qq,Lee-2,Groote2014}. In Refs.\cite{Nielsen2005,Wang-nonet}, the scalar nonet mesons
below $1\,\rm{ GeV}$ are taken to be  the tetraquark states consist of scalar diquark pairs  and studied with the QCD sum rules by carrying out the operator product expansion up to the vacuum condensates of dimension 6. In Ref.\cite{Lee2006}, Lee carries out the operator product expansion by including the vacuum condensates up
to dimension 8, and observes   no evidence of the couplings of the tetraquark currents to the
light scalar   nonet mesons. In Ref.\cite{Chen-Zhu}, Chen, Hosaka and Zhu study the light scalar tetraquark states with the QCD sum rules in a systematic way.
In Ref.\cite{tetra-qq}, Sugiyama et al study the non-singlet scalar mesons
$a_0(980)$ and $\kappa_0(800)$ as the two-quark-tetraquark  mixed states with the QCD sum rules, and observe that  the tetraquark  currents predict lower masses than the two-quark currents, and  the tetraquark states occupy about $(70-90)\%$ of the lowest mass states.

 In this article, we assume that the scalar nonet  mesons below  $1\,\rm{GeV}$ are the two-quark-tetraquark mixed states  and study their properties
 with the  QCD sum rules in a systematic way by taking into account the vacuum condensates up to dimension 10  and the  $\mathcal{O}(\alpha_s)$ corrections to the dimension zero  terms in the QCD spectral densities in the operator product expansion.

The article is arranged as follows:  we derive the QCD sum rules for the scalar nonet mesons  in Sect.2;
in Sect.3, we present the numerical results and discussions; and Sect.4 is reserved for our
conclusions.

\section{ The scalar  nonet mesons  with the  QCD Sum Rules}

In the  scenario of conventional two-quark  states, the structures of the scalar nonet
mesons in the ideal mixing limit can be symbolically written as
\begin{eqnarray}
f_0(500)= \frac{\bar{u}u+\bar{d}d}{\sqrt{2}}\, ,\;&&f_0(980)= \bar{s}s\, , \nonumber\\
a_0^-(980)=d\bar{u},\;&&a_0^0(980)={u\bar{u}-d\bar{d}\over\sqrt{2}}\, ,\;\;\;\;\;\;a_0^+(980)=u\bar{d}\, ,\nonumber\\
\kappa_0^+(800)=u \bar{s}\, ,\;&&\kappa_0^0(800)= d \bar{s}\, ,\;\;\;\;\;\;\bar{\kappa}_0^0(800)= s \bar{d}\, ,
\;\;\;\;\;\;\kappa_0^-(800)= s\bar{u}  \, .
\end{eqnarray}

In the  scenario of tetraquark  states, the structures of the scalar nonet
mesons in the ideal mixing limit can be symbolically written as
\cite{Close2002,ReviewJaffe,ReviewAmsler2}
\begin{eqnarray}
f_0(500)=ud\bar{u}\bar{d}\, ,\;&&f_0(980)={us\bar{u}\bar{s}+ds\bar{d}\bar{s}\over\sqrt{2}}\, , \nonumber\\
a_0^-(980)=ds\bar{u}\bar{s},\;&&a_0^0(980)={us\bar{u}\bar{s}-ds\bar{d}\bar{s}\over\sqrt{2}}\, ,\;\;\;\;\;\;a_0^+(980)=us\bar{d}\bar{s}\, ,\nonumber\\
\kappa_0^+(800)=ud\bar{d}\bar{s}\, ,\;&&\kappa_0^0(800)=ud\bar{u}\bar{s}\, ,\;\;\;\;\;\;\bar{\kappa}_0^0(800)=us\bar{u}\bar{d}\, ,
\;\;\;\;\;\;\kappa_0^-(800)=ds\bar{u}\bar{d} \, .
\end{eqnarray}
If we take   the diquarks and
antidiquarks as the basic constituents,  the two isoscalar states $\bar
u\bar d u d$ and $\bar s s\frac{\bar u u+\bar d d}{\sqrt{2}}$ mix
ideally, the $\bar s s\frac{\bar u u+\bar d d}{\sqrt{2}}$
 degenerates with the isovector states  $\bar s s\bar d u$,
$\bar s s\frac{\bar u u-\bar d d}{\sqrt{2}}$ and $\bar s s\bar u
d$ naturally.  The mass spectrum is inverted compare to  the traditional ${\bar q} q $
mesons.  The lightest state is the
non-strange isosinglet, the heaviest states are the
degenerate isosinglet and isovector states with hidden $\bar s s$ pairs,
  the four strange states lie in between.

In this article, we take the   scalar nonet mesons to be  the two-quark-tetraquark  mixed states, and write down the  two-point correlation functions $\Pi_S(p)$,
\begin{eqnarray}
\Pi_S(p^2)&=&i\int d^4x ~e^{ip\cdot x}\langle 0|T\left\{J_S(x){J_S}^\dagger(0)\right\}|0\rangle \, , \\
 J_{S}(x)&=&\cos\theta_S J_{S}^4(x)+\sin\theta_S J_{S}^2(x) \, ,
\end{eqnarray}
where $S=f_0(980),\,a_0^0(980),\,\kappa_0^+(800),\,f_0(500)$, and
\begin{eqnarray}
J^4_{f_0(980)}(x)&=&\frac{\epsilon^{ijk}\epsilon^{imn}}{\sqrt{2}}\left\{u_j^T(x)C\gamma_5s_k(x)\, \bar{u}_m(x)\gamma_5C\bar{s}_n^T(x)+d_j^T(x)C\gamma_5s_k(x)\,\bar{d}_m(x)\gamma_5C\bar{s}_n^T(x)\right\}\, , \nonumber \\
J^2_{f_0(980)}(x)&=&-\frac{\langle\bar{q}q\rangle}{3\sqrt{2}} \bar{s}(x)s(x)\, , \\
J^4_{a_0^0(980)}(x)&=&\frac{\epsilon^{ijk}\epsilon^{imn}}{\sqrt{2}}\left\{u_j^T(x)C\gamma_5s_k(x)\,\bar{u}_m(x)\gamma_5C\bar{s}_n^T(x)-d_j^T(x)C\gamma_5s_k(x)\,\bar{d}_m(x)\gamma_5C\bar{s}_n^T(x)\right\}\, , \nonumber\\
J^2_{a_0^0(980)}(x)&=& -\frac{\langle\bar{s}s\rangle}{6} \frac{\bar{u}(x)u(x)-\bar{d}(x)d(x)}{\sqrt{2}}\, ,  \\
J^4_{\kappa_0^+(800)}(x)&=&\epsilon^{ijk}\epsilon^{imn}\, u_j^T(x)C\gamma_5d_k(x)\, \bar{s}_m(x)\gamma_5C\bar{d}_n^T(x)\, ,\nonumber \\
J^2_{\kappa_0^+(800)}(x)&=&-\frac{\langle\bar{q}q\rangle}{6} \bar{s}(x)u(x)\, ,\\
J^4_{f_0(500)}(x)&=&\epsilon^{ijk}\epsilon^{imn}\, u_j^T(x)C\gamma_5d_k(x)\, \bar{u}_m(x)\gamma_5C\bar{d}_n^T(x)\, ,\nonumber \\
J^2_{f_0(500)}(x)&=&-\frac{\langle\bar{q}q\rangle}{3\sqrt{2}}\frac{\bar{u}(x)u(x)+\bar{d}(x)d(x)}{\sqrt{2}} \, ,
\end{eqnarray}
the currents $J^4_S(x)$ and $J^2_S(x)$ are tetraquark and two-quark operators, respectively, and couple potentially to the tetraquark and two-quark components  of the scalar nonet mesons, respectively, the $\theta_S$ are the mixing angles. In the currents $J^4_S(x)$, the $i,~j,~k,~...$ are color indices and $C$ is the charge
conjugation matrix,  the $ \epsilon^{ijk}
u_j^T(x)C\gamma_5 d_k(x)$, $\epsilon^{ijk} u_j^T(x)C\gamma_5
s_k(x)$, $\epsilon^{ijk} d_j^T(x)C\gamma_5 s_k(x) $ represent the
scalar   diquarks in color anti-triplet, the corresponding antidiquarks can be obtained by charge conjugation.   The one-gluon exchange
force and the instanton induced force can result in  significant
attractions between the quarks in the scalar diquark channels
\cite{ReviewJaffe,Instanton}.

In the following, we  perform Fierz re-arrangement to the currents $J^4_{f_0(980)}$ and $J^4_{a^0_0(980)}$ both in the color and Dirac-spinor  spaces to obtain the  result,
\begin{eqnarray}
J^4_{f_0(980)}&=&\frac{1}{4}\left\{\,-\bar{s} s \frac{\bar{u} u+\bar{d} d}{\sqrt{2}}+\bar{s}i\gamma_5 s\frac{\bar{u}i\gamma_5 u+\bar{d}i\gamma_5 d}{\sqrt{2}}-\bar{s} \gamma^\mu s\frac{\bar{u}\gamma_\mu u+\bar{d}\gamma_\mu d}{\sqrt{2}}-\bar{s} \gamma^\mu\gamma_5 s\frac{\bar{u}\gamma_\mu\gamma_5 u+\bar{d}\gamma_\mu \gamma_5 d}{\sqrt{2}}\right. \nonumber\\
&&+\frac{1}{2}\bar{s}\sigma_{\mu\nu} s \frac{\bar{u}\sigma^{\mu\nu} u+\bar{d}\sigma^{\mu\nu} d}{\sqrt{2}} +\frac{\bar{s} u\,\bar{u} s+\bar{s} d\,\bar{d} s}{\sqrt{2}}-\frac{\bar{s}i\gamma_5 u\,\bar{u}i\gamma_5 s+\bar{s}i\gamma_5 d\,\bar{d}i\gamma_5 s}{\sqrt{2}}\nonumber\\
&&+\frac{\bar{s} \gamma^\mu u\,\bar{u}\gamma_\mu s+\bar{s} \gamma^\mu d\,\bar{d}\gamma_\mu s}{\sqrt{2}}+\frac{\bar{s} \gamma^\mu\gamma_5 u\,\bar{u}\gamma_\mu\gamma_5 s+\bar{s} \gamma^\mu\gamma_5 d\,\bar{d}\gamma_\mu\gamma_5 s}{\sqrt{2}}\nonumber\\
&&\left.-\frac{1}{2}\frac{\bar{s}\sigma_{\mu\nu} u\,\bar{u}\sigma^{\mu\nu} s+\bar{s}\sigma_{\mu\nu} d\,\bar{d}\sigma^{\mu\nu} s}{\sqrt{2}}  \,\right\} \, ,
\end{eqnarray}
\begin{eqnarray}
J^4_{a_0^0(980)}&=&\frac{1}{4}\left\{\,-\bar{s} s \frac{\bar{u} u-\bar{d} d}{\sqrt{2}}+\bar{s}i\gamma_5 s\frac{\bar{u}i\gamma_5 u-\bar{d}i\gamma_5 d}{\sqrt{2}}-\bar{s} \gamma^\mu s\frac{\bar{u}\gamma_\mu u-\bar{d}\gamma_\mu d}{\sqrt{2}}-\bar{s} \gamma^\mu\gamma_5 s\frac{\bar{u}\gamma_\mu\gamma_5 u-\bar{d}\gamma_\mu \gamma_5 d}{\sqrt{2}}\right. \nonumber\\
&&+\frac{1}{2}\bar{s}\sigma_{\mu\nu} s \frac{\bar{u}\sigma^{\mu\nu} u-\bar{d}\sigma^{\mu\nu} d}{\sqrt{2}} +\frac{\bar{s} u\,\bar{u} s-\bar{s} d\,\bar{d} s}{\sqrt{2}}-\frac{\bar{s}i\gamma_5 u\,\bar{u}i\gamma_5 s-\bar{s}i\gamma_5 d\,\bar{d}i\gamma_5 s}{\sqrt{2}}\nonumber\\
&&+\frac{\bar{s} \gamma^\mu u\,\bar{u}\gamma_\mu s-\bar{s} \gamma^\mu d\,\bar{d}\gamma_\mu s}{\sqrt{2}}+\frac{\bar{s} \gamma^\mu\gamma_5 u\,\bar{u}\gamma_\mu\gamma_5 s-\bar{s} \gamma^\mu\gamma_5 d\,\bar{d}\gamma_\mu\gamma_5 s}{\sqrt{2}}\nonumber\\
&&\left.-\frac{1}{2}\frac{\bar{s}\sigma_{\mu\nu} u\,\bar{u}\sigma^{\mu\nu} s-\bar{s}\sigma_{\mu\nu} d\,\bar{d}\sigma^{\mu\nu} s}{\sqrt{2}}  \,\right\} \, ,
\end{eqnarray}
 some components  couple potentially  to the meson pairs $\pi\pi$, $K\bar{K}$, $\eta \pi$, the strong decays $f_0(980)\to \pi\pi$, $K\bar{K}$ and $a^0_0(980)\to \eta\pi$, $K\bar{K}$ are Okubo-Zweig-Iizuka super-allowed, which can also be used to study the radiative decays $\phi(1020)\to f_0(980)\gamma$ and $\phi(1020)\to a_0^0(980)\gamma$ through the  virtual   $K\bar{K}$ loops. So it is reasonable to  assume that the nonet scalar mesons  below $1\,\rm{ GeV}$ have some tetraquark constituents.

 The tetraquark operator $J^4_S(x)$ contains a hidden $\bar{q}q$ component with $q=u$, $d$ or $s$.  If we contract  the corresponding quark pair in the currents $J^4_{S}(x)$ and substitute it by  the quark condensate \footnote{ For example,
 \begin{eqnarray}
 J^4_{f_0(980)}&=&\frac{\epsilon^{ijk}\epsilon^{imn}}{\sqrt{2}}\left\{ \left[C\gamma_5\right]_{\alpha\beta}\left[\gamma_5C\right]_{\lambda\tau} u^j_\alpha s^k_\beta \bar{u}^m_\lambda \bar{s}^n_\tau +\left[C\gamma_5\right]_{\alpha\beta}\left[\gamma_5C\right]_{\lambda\tau} d^j_\alpha s^k_\beta \bar{d}^m_\lambda \bar{s}^n_\tau \right\} \nonumber\\
 &=&\frac{\epsilon^{ijk}\epsilon^{imn}}{\sqrt{2}}\left\{- \left[C\gamma_5\right]_{\alpha\beta}\left[\gamma_5C\right]_{\lambda\tau}   \bar{u}^m_\lambda u^j_\alpha \bar{s}^n_\tau s^k_\beta - \left[C\gamma_5\right]_{\alpha\beta}\left[\gamma_5C\right]_{\lambda\tau}   \bar{d}^m_\lambda d^j_\alpha \bar{s}^n_\tau s^k_\beta \right\} \nonumber\\
  &\to&\frac{\epsilon^{ijk}\epsilon^{imn}}{\sqrt{2}}\left\{- \left[C\gamma_5\right]_{\alpha\beta}\left[\gamma_5C\right]_{\lambda\tau}   \frac{\delta_{jm}\delta_{\lambda\alpha}}{12} \langle \bar{u}u\rangle \frac{\delta_{nk}\delta_{\tau\beta}}{12} \bar{s}  s  - \left[C\gamma_5\right]_{\alpha\beta}\left[\gamma_5C\right]_{\lambda\tau}   \frac{\delta_{jm}\delta_{\lambda\alpha}}{12} \langle \bar{d}d\rangle
  \frac{\delta_{nk}\delta_{\tau\beta}}{12} \bar{s}  s \right\} \nonumber\\
  &=&-\frac{\langle\bar{u}u\rangle+\langle\bar{d}d\rangle}{24\sqrt{2}} \,Tr\left\{ \left[C\gamma_5\right]\left[\gamma_5C\right]^T \right\} \bar{s}s=-\frac{\langle \bar{q}q\rangle}{3\sqrt{2}}\,\bar{s}s=J^2_{f_0(980)}\, , \nonumber
 \end{eqnarray}
 where the $\alpha$, $\beta$, $\lambda$ and $\tau$ are Dirac spinor indexes.},  then
\begin{eqnarray}
J^4_{f_0(980)}(x)&\to&J^2_{f_0(980)}(x)\, , \nonumber \\
J^4_{a_0^0(980)}(x)& \to & J^2_{a_0^0(980)}(x)\, , \nonumber \\
J^4_{\kappa_0^+(800)}(x)& \to& J^2_{\kappa_0^+(800)}(x) \, , \nonumber\\
J^4_{f_0(500)}(x)& \to & J^2_{f_0(500)}(x)\, .
\end{eqnarray}
The  contracted parts appear as the normalization factors $-\frac{\langle\bar{q}q\rangle}{3\sqrt{2}}$, $-\frac{\langle\bar{s}s\rangle}{6}$, $-\frac{\langle\bar{q}q\rangle}{6}$ and  $-\frac{\langle\bar{q}q\rangle}{3\sqrt{2}}$ in the currents $J^2_{f_0(980)}(x)$, $J^2_{a_0(980)}(x)$,  $J^2_{\kappa_0(800)}(x)$ and $J^2_{f_0(500)}(x)$, respectively.

 We  insert  a complete set  of intermediate   states  with the same quantum numbers as the current operators  $J_S(x)$
 satisfying the unitarity   principle into the correlation functions  $\Pi_S(p^2)$  to obtain the hadronic representation \cite{SVZ79,PRT85}. After isolating the
ground state contributions from the pole terms of the scalar nonet
mesons, we get the  result,
\begin{eqnarray}
\Pi_S(p^2)=\frac{\lambda_S^2}{m_S^{2}-p^2}+\cdots \, ,
\end{eqnarray}
where we have used the definitions $ \langle 0 | J_S(0)|S\rangle =\lambda_S$ for the pole residues.

The correlation functions can be re-written as
\begin{eqnarray}
\Pi_S(p^2)&=&\cos^2\theta\, \Pi_S^{44}(p^2) +\sin\theta \cos\theta \, \Pi_S^{42}(p^2)+\sin\theta \cos\theta \, \Pi_S^{24}(p^2)+\sin^2\theta  \, \Pi_S^{22}(p^2) \, ,\nonumber \\
\Pi^{mn}_S(p^2)&=&i\int d^4x ~e^{ip\cdot x}\langle 0|T\left\{J_S^{m}(x){J_S^n}^{\dagger}(0)\right\}|0\rangle \, ,
\end{eqnarray}
where $m,n=2,4$.  We can prove that $\Pi^{mn}_S(p^2)=\Pi^{nm}_S(p^2)$ with the replacements $x \to -x$ and $p \to -p$ for $m\neq n$.

In the following, we briefly outline  the operator product expansion for the correlation functions $\Pi^{mn}_S(p^2)$   in perturbative QCD. Firstly,  we contract the $u$, $d$ and $s$ quark fields in the correlation functions
$\Pi^{mn}_S(p^2)$    with Wick theorem, and obtain the results:
\begin{eqnarray}
\Pi^{44}_{f_0/a_0(980)}(p^2)&=& \frac{i}{2}\, \varepsilon^{ijk}\varepsilon^{i^\prime j^\prime k^\prime}\varepsilon^{imn}\varepsilon^{i^\prime m^\prime n^\prime}\int d^4x ~e^{ip\cdot x}\nonumber\\
&&\left\{ {\rm Tr} \left[\gamma_5 S_{kk^\prime}(x)\gamma_5 CU^{T}_{jj^\prime}(x) C\right]{\rm Tr} \left[\gamma_5 S_{n^\prime n}(-x)\gamma_5 CU^{T}_{m^\prime m}(-x) C\right] \right. \nonumber\\
&&\left.+ {\rm Tr} \left[\gamma_5 S_{kk^\prime}(x)\gamma_5 CD^{T}_{jj^\prime}(x) C\right]{\rm Tr} \left[\gamma_5 S_{n^\prime n}(-x)\gamma_5 CD^{T}_{m^\prime m}(-x) C\right] \right\} \, ,\nonumber\\
\Pi^{44}_{\kappa_0(800)}(p^2)&=& i\, \varepsilon^{ijk}\varepsilon^{i^\prime j^\prime k^\prime}\varepsilon^{imn}\varepsilon^{i^\prime m^\prime n^\prime}\int d^4x ~e^{ip\cdot x}\nonumber\\
&& {\rm Tr} \left[\gamma_5 D_{kk^\prime}(x)\gamma_5 CU^{T}_{jj^\prime}(x) C\right]{\rm Tr} \left[\gamma_5 D_{n^\prime n}(-x)\gamma_5 CS^{T}_{m^\prime m}(-x) C\right] \, ,\nonumber\\
\Pi^{44}_{f_0(500)}(p^2)&=& i\, \varepsilon^{ijk}\varepsilon^{i^\prime j^\prime k^\prime}\varepsilon^{imn}\varepsilon^{i^\prime m^\prime n^\prime}\int d^4x ~e^{ip\cdot x}\nonumber\\
&& {\rm Tr} \left[\gamma_5 D_{kk^\prime}(x)\gamma_5 CU^{T}_{jj^\prime}(x) C\right]{\rm Tr} \left[\gamma_5 D_{n^\prime n}(-x)\gamma_5 CU^{T}_{m^\prime m}(-x) C\right] \, ,
\end{eqnarray}

\begin{eqnarray}
\Pi_{f_0(980)}^{42}(p^2)&=&-\frac{\langle \bar{q}q\rangle^2}{18}\, i \, \int d^4x ~e^{ip\cdot x} {\rm Tr} \left[S_{jk}(x)S_{kj}(-x) \right]\nonumber\\
&&+ \frac{\langle \bar{q}  q\rangle}{24} \varepsilon^{ijk}\varepsilon^{imn} \langle \bar{q}_{m} \sigma_{\mu\nu} q_j\rangle\, i\,\int d^4x ~e^{ip\cdot x} {\rm Tr} \left[S_{ka}(x)S_{an}(-x)\sigma^{\mu\nu} \right]\, ,\nonumber\\
\Pi_{a_0(980)}^{42}(p^2)&=&-\frac{\langle \bar{s}s\rangle^2}{72}\, i \, \int d^4x ~e^{ip\cdot x} \left\{{\rm Tr} \left[U_{jk}(x)U_{kj}(-x)\right]+
{\rm Tr}\left[ D_{jk}(x)D_{kj}(-x)\right] \right\}\nonumber\\
&&+ \frac{\langle \bar{s}  s\rangle}{96} \varepsilon^{ijk}\varepsilon^{imn} \langle \bar{s}_{n} \sigma_{\mu\nu} s_k\rangle\, i\,\int d^4x ~e^{ip\cdot x}\nonumber\\
&& \left\{ {\rm Tr} \left[U_{ja}(x)U_{am}(-x)\sigma^{\mu\nu} \right]+{\rm Tr} \left[D_{ja}(x)D_{am}(-x)\sigma^{\mu\nu} \right]\right\} \, ,\nonumber\\
\Pi_{\kappa_0(800)}^{42}(p^2)&=&-\frac{\langle \bar{q}q\rangle^2}{36}\, i \, \int d^4x ~e^{ip\cdot x} {\rm Tr} \left[U_{jk}(x)S_{kj}(-x) \right]\nonumber\\
&&+ \frac{\langle \bar{q}  q\rangle}{48} \varepsilon^{ijk}\varepsilon^{imn} \langle \bar{q}_{n} \sigma_{\mu\nu} q_k\rangle\, i\,\int d^4x ~e^{ip\cdot x} {\rm Tr} \left[U_{ja}(x)S_{am}(-x)\sigma^{\mu\nu} \right]\, ,\nonumber\\
\Pi_{f_0(500)}^{42}(p^2)&=&-\frac{\langle \bar{q}q\rangle^2}{36}\, i \, \int d^4x ~e^{ip\cdot x} \left\{ {\rm Tr} \left[U_{jk}(x)U_{kj}(-x) \right]+ {\rm Tr} \left[D_{jk}(x)D_{kj}(-x) \right]\right\}\nonumber\\
&&+ \frac{\langle \bar{q}  q\rangle}{48} \varepsilon^{ijk}\varepsilon^{imn} \langle \bar{q}_{n} \sigma_{\mu\nu} q_k\rangle\, i\,\int d^4x ~e^{ip\cdot x} \nonumber\\
&&\left\{{\rm Tr} \left[U_{ja}(x)U_{am}(-x)\sigma^{\mu\nu} \right]+{\rm Tr} \left[D_{ja}(x)D_{am}(-x)\sigma^{\mu\nu} \right]\right\}\, ,
\end{eqnarray}

\begin{eqnarray}
\Pi_{f_0(980)}^{24}(p^2)&=&\Pi_{f_0(980)}^{42}(p^2)\, , \nonumber\\
\Pi_{a_0(980)}^{24}(p^2)&=&\Pi_{a_0(980)}^{42}(p^2)\, , \nonumber\\
\Pi_{\kappa_0(800)}^{24}(p^2)&=&\Pi_{\kappa_0(800)}^{42}(p^2)\, , \nonumber\\
\Pi_{f_0(500)}^{24}(p^2)&=&\Pi_{f_0(500)}^{42}(p^2)\, ,
\end{eqnarray}

\begin{eqnarray}
\Pi_{f_0(980)}^{22}(p^2)&=&-\frac{\langle \bar{q}q\rangle^2}{18}\, i \, \int d^4x ~e^{ip\cdot x} {\rm Tr} \left[S_{jk}(x)S_{kj}(-x) \right]\, , \nonumber\\
\nonumber\\
\Pi_{a_0(980)}^{22}(p^2)&=&-\frac{\langle \bar{s}s\rangle^2}{72}\, i \, \int d^4x ~e^{ip\cdot x} \left\{{\rm Tr} \left[U_{jk}(x)U_{kj}(-x)\right]+
{\rm Tr}\left[ D_{jk}(x)D_{kj}(-x)\right] \right\}\, ,\nonumber\\
\Pi_{\kappa_0(800)}^{22}(p^2)&=&-\frac{\langle \bar{q}q\rangle^2}{36}\, i \, \int d^4x ~e^{ip\cdot x} {\rm Tr} \left[U_{jk}(x)S_{kj}(-x) \right]\, ,\nonumber\\
\Pi_{f_0(500)}^{22}(p^2)&=&-\frac{\langle \bar{q}q\rangle^2}{36}\, i \, \int d^4x ~e^{ip\cdot x} \left\{ {\rm Tr} \left[U_{jk}(x)U_{kj}(-x) \right]+ {\rm Tr} \left[D_{jk}(x)D_{kj}(-x) \right]\right\}\, ,
\end{eqnarray}
where
\begin{eqnarray}
U_{ij}(x)&=& \frac{i\delta_{ij}\!\not\!{x}}{ 2\pi^2x^4}
-\frac{\delta_{ij}m_q}{4\pi^2x^2}-\frac{\delta_{ij}\langle
\bar{q}q\rangle}{12} +\frac{i\delta_{ij}\!\not\!{x}m_q
\langle\bar{q}q\rangle}{48}-\frac{\delta_{ij}x^2\langle \bar{q}g_s\sigma Gq\rangle}{192}+\frac{i\delta_{ij}x^2\!\not\!{x} m_q\langle \bar{q}g_s\sigma
 Gq\rangle }{1152}\nonumber\\
&& -\frac{ig_s G^{a}_{\alpha\beta}t^a_{ij}(\!\not\!{x}
\sigma^{\alpha\beta}+\sigma^{\alpha\beta} \!\not\!{x})}{32\pi^2x^2}  -\frac{1}{8}\langle\bar{q}_j\sigma^{\mu\nu}q_i \rangle \sigma_{\mu\nu}+\cdots \, ,\nonumber\\
D_{ij}(x)&=&U_{ij}(x) \, , \nonumber\\
S_{ij}(x)&=& \frac{i\delta_{ij}\!\not\!{x}}{ 2\pi^2x^4}
-\frac{\delta_{ij}m_s}{4\pi^2x^2}-\frac{\delta_{ij}\langle
\bar{s}s\rangle}{12} +\frac{i\delta_{ij}\!\not\!{x}m_s
\langle\bar{s}s\rangle}{48}-\frac{\delta_{ij}x^2\langle \bar{s}g_s\sigma Gs\rangle}{192}+\frac{i\delta_{ij}x^2\!\not\!{x} m_s\langle \bar{s}g_s\sigma
 Gs\rangle }{1152}\nonumber\\
&& -\frac{ig_s G^{a}_{\alpha\beta}t^a_{ij}(\!\not\!{x}
\sigma^{\alpha\beta}+\sigma^{\alpha\beta} \!\not\!{x})}{32\pi^2x^2}  -\frac{1}{8}\langle\bar{s}_j\sigma^{\mu\nu}s_i \rangle \sigma_{\mu\nu}+\cdots \, ,
\end{eqnarray}
where $q=u,d$ \cite{PRT85}.  We take the assumption of vacuum saturation for the  higher dimension vacuum condensates and factorize the higher dimension vacuum condensates into lower dimension vacuum condensates \cite{SVZ79}, for example, $\langle \bar{q}q\bar{q}q\rangle\sim \langle \bar{q}q\rangle\langle \bar{q}q\rangle$, $\langle \bar{q}q\bar{q}g_s\sigma Gq\rangle\sim \langle \bar{q}q\rangle\langle \bar{q}g_s \sigma Gq\rangle$, where $q=u,d,s$. Factorization works well in  large $N_c$ limit,  in reality, $N_c=3$, some  (not much) ambiguities maybe originate  from
the vacuum saturation assumption.

In Fig.1, we show the Feynman diagrams containing the $\bar{q}q$ annihilations   accounting for the mixing of different
Fock states. The quark-pair annihilations are substituted
by the condensates $\langle \bar{q}q\rangle\langle \bar{q}^\prime q^\prime\rangle$ and $\langle \bar{q}q\rangle\langle \bar{q}^\prime g_s\sigma Gq^\prime\rangle$ as there are normalization factors $\langle \bar{q}q\rangle$ in the  interpolating currents $J^2_S(x)$. The perturbative
part of the quark-pair annihilations must disappear as only the terms $\langle
\bar{q}q\rangle$ and $\langle\bar{q}_j\sigma^{\mu\nu}q_i \rangle$ in the full quark propagators $U_{ij}(x)$, $D_{ij}(x)$ and $S_{ij}(x)$
survive in the limit $x \to 0$, where $q=u,d,s$.

In Eq.(18), we retain the terms $\langle\bar{q}_j\sigma_{\mu\nu}q_i \rangle$ and $\langle\bar{s}_j\sigma_{\mu\nu}s_i \rangle$  come from the Fierz re-arrangement of the $\langle q_i \bar{q}_j\rangle$ and $\langle s_i \bar{s}_j\rangle$ to  absorb the gluons  emitted from other  quark lines to form $\langle\bar{q}_j g_s G^a_{\alpha\beta} t^a_{mn}\sigma_{\mu\nu} q_i \rangle$ and $\langle\bar{s}_j g_s G^a_{\alpha\beta} t^a_{mn}\sigma_{\mu\nu} s_i \rangle$  to extract the mixed condensates   $\langle\bar{q}g_s\sigma G q\rangle$ and $\langle\bar{s}g_s\sigma G s\rangle$.   Some terms involving the mixed condensates  $\langle\bar{q}g_s\sigma G q\rangle$ and $\langle\bar{s}g_s\sigma G s\rangle$ appear and play an important role in the QCD sum rules, see the second Feynman diagram shown in Fig.1 and the first two Feynman diagrams shown in Fig.2.

\begin{figure}
 \centering
 \includegraphics[totalheight=3cm,width=12cm]{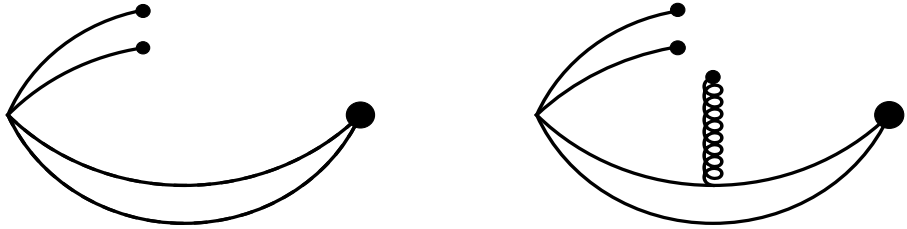}
    \caption{The    Feynman diagrams  contribute to the condensates $\langle \bar{q}q\rangle\langle \bar{q}^\prime q^\prime\rangle$ and $\langle \bar{q}q\rangle\langle \bar{q}^\prime g_s\sigma Gq^\prime\rangle$ in the correlation functions $\Pi_S^{42}(p^2)$, where $q, q^\prime=u,d,s$ and $S=f_0(980)$, $a_0(980)$, $\kappa_0(800)$, $f_0(500)$, the large $\bullet$ denotes the normalization factors $\langle \bar{q}q\rangle$ in the currents $J^2_S(0)$. Other diagrams obtained  by interchanging of the  quark lines are implied.  }
\end{figure}

\begin{figure}
 \centering
 \includegraphics[totalheight=3cm,width=12cm]{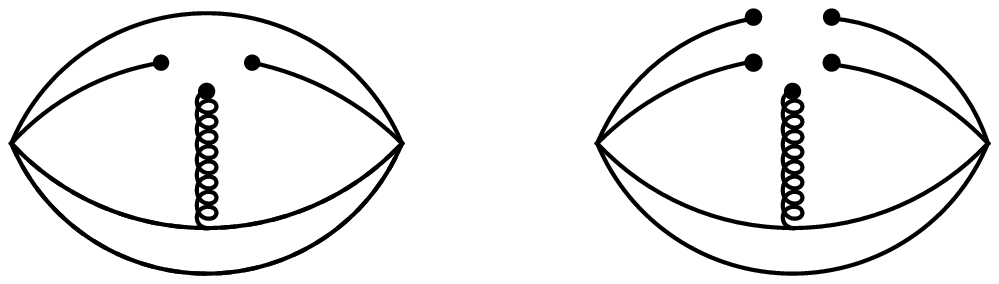}
 \vglue+4mm
 \includegraphics[totalheight=3cm,width=12cm]{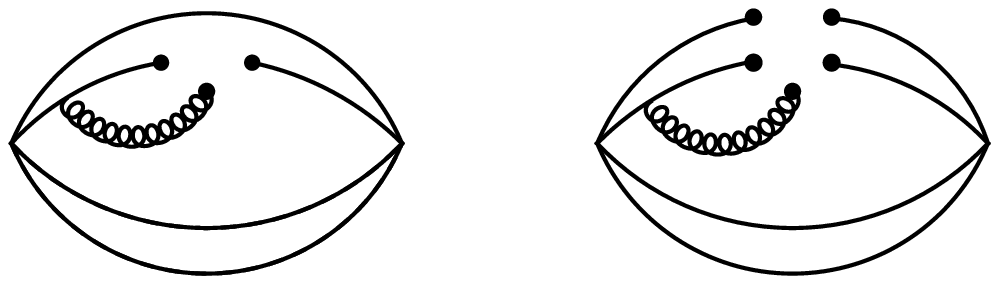}
     \caption{The    Feynman diagrams  contribute to the condensates $\langle \bar{q}g_s\sigma Gq\rangle$ and $\langle \bar{q}q\rangle\langle \bar{q}^\prime g_s\sigma Gq^\prime\rangle$ in the correlation functions $\Pi_S^{44}(p^2)$, where $q, q^\prime=u,d,s$ and $S=f_0(980)$, $a_0(980)$, $\kappa_0(800)$, $f_0(500)$. Other diagrams obtained  by interchanging of the  quark lines are implied.  }
\end{figure}

Then we compute  the integrals  in the coordinate  space to obtain the correlation functions $\Pi_S(p^2)$,   therefore the QCD spectral densities $\rho_{S}(s)$  at the quark level through the dispersion  relation,
\begin{eqnarray}
\rho_S(s)&=&\frac{{\rm Im}\Pi(s)}{\pi} \, .
\end{eqnarray}
In this article, we approximate the continuum contributions by
\begin{eqnarray}
\int_{s^0_S}^\infty ds \,\rho_S(s)\,\exp\left({-\frac{s}{M^2}}\right) \, ,
\end{eqnarray}
which contain both perturbative and non-perturbative contributions, we use the $s_S^0$ to denote the continuum threshold
parameters. For the conventional two-quark scalar  mesons, only perturbative
 contributions survive in such integrals, see Eqs.(26-27), Eq.(30) and Eq.(33).

In this article, we carry out the operator product expansion by including the  vacuum condensates up to dimension 10.
 The condensates $\langle g_s^3 GGG\rangle$, $\langle \frac{\alpha_s GG}{\pi}\rangle^2$,
 $\langle \frac{\alpha_s GG}{\pi}\rangle\langle \bar{q} g_s \sigma Gq\rangle$ have the dimensions 6, 8, 9 respectively,  but they are   the vacuum expectations
of the operators of the order    $\mathcal{O}( \alpha_s^{3/2})$, $\mathcal{O}(\alpha_s^2)$, $\mathcal{O}( \alpha_s^{3/2})$ respectively, their values  are very small
and discarded.  We take
the truncations $n\leq 10$ and $k\leq 1$,
the operators of the orders $\mathcal{O}( \alpha_s^{k})$ with $k> 1$ are  discarded.  Furthermore, we take into account the
$\mathcal{O}( \alpha_s)$ corrections to the perturbative terms, which were  calculated recently \cite{Groote2014}. As there are normalization factors  $\langle \bar{q}q\rangle^2$ in the correlation functions $\Pi_S^{22}(p)$ ,  we count those  perturbative terms as of the order $\langle \bar{q}q\rangle^2$, and truncate the operator product expansion to the order $\langle \bar{q}q\rangle^2\langle\bar{q}^\prime q^\prime\rangle$, where $q,q^\prime=u,d,s$.

  Once  the analytical  QCD spectral densities are obtained,
  then we can take the quark-hadron duality  below the continuum thresholds
$s_S^0$ and perform the Borel transformation with respect to the
variable $P^2=-p^2$, finally we obtain  the  QCD sum rules,
\begin{eqnarray}
\lambda^{2}_{S} \exp\left({-\frac{m_{S}^{2}}{M^2}}\right)&=&\int_0^{s_S^0}ds \,\rho_S(s)\,\exp\left({-\frac{s}{M^2}}\right) \, ,
\end{eqnarray}

\begin{eqnarray}
\rho_S(s)&=&\cos^2\theta_S\,\rho^{44}_S(s)+2\sin\theta_S \cos\theta_S \,\rho^{42}_S(s)+\sin^2\theta_S \,\rho^{22}_S(s) \, ,
\end{eqnarray}

\begin{eqnarray}
\rho^{44}_{f_0/a_0(980)}&=&\frac{s^4}{61440\pi^6}\left\{1+\frac{\alpha_s}{\pi}\left( \frac{57}{5}+2\log \frac{\mu^2}{s}\right) \right\}+\frac{(m_q-2m_s)\langle\bar{q}q\rangle+(m_s-2m_q)\langle\bar{s}s\rangle}{192\pi^4}s^2
 \nonumber\\
&&+\frac{(3m_s-m_q)\langle\bar{q}g_s\sigma Gq\rangle+(3m_q-m_s)\langle\bar{s}g_s\sigma Gs\rangle}{192\pi^4}s+\frac{\langle\bar{q}q\rangle\langle\bar{s}s\rangle}{12\pi^2}s\nonumber\\
&&-\frac{\langle\bar{q}q\rangle\langle\bar{s}g_s\sigma Gs\rangle+\langle\bar{s}s\rangle\langle\bar{q}g_s\sigma Gq\rangle}{24\pi^2}
+\frac{\langle\bar{q}g_s\sigma Gq\rangle\langle\bar{s}g_s\sigma Gs\rangle}{96\pi^2}\delta(s)\nonumber\\
&&-\frac{(2m_q-m_s)\langle\bar{q}q\rangle\langle\bar{s}s\rangle^2+(2m_s-m_q)\langle\bar{s}s\rangle\langle\bar{q}q\rangle^2}{9}\delta(s)+\frac{s^2}{1536\pi^4}\langle\frac{\alpha_sGG}{\pi}\rangle\nonumber\\
&&-\frac{m_s\langle\bar{q}q\rangle+m_q\langle\bar{s}s\rangle}{72\pi^2}\langle\frac{\alpha_sGG}{\pi}\rangle+\frac{m_q\langle\bar{q}q\rangle+m_s\langle\bar{s}s\rangle}{192\pi^2}\langle\frac{\alpha_sGG}{\pi}\rangle\nonumber\\
&&+\frac{5}{216}\langle\bar{q}q\rangle\langle\bar{s}s\rangle\langle\frac{\alpha_sGG}{\pi}\rangle\delta(s)\, ,\\
\rho^{42}_{f_0(980)}&=&\frac{\langle\bar{q}q\rangle^2}{144}\left\{\frac{3}{\pi^2}s+\langle\frac{\alpha_sGG}{\pi}\rangle\delta(s)+24m_s\langle\bar{s}s\rangle\delta(s)\right\}
+\frac{\langle\bar{q}q\rangle\langle\bar{q}g_s\sigma Gq\rangle}{96\pi^2}\, , \\
\rho^{42}_{a_0(980)}&=&\frac{\langle\bar{s}s\rangle^2}{288}\left\{\frac{3}{\pi^2}s+\langle\frac{\alpha_sGG}{\pi}\rangle\delta(s)+24m_q\langle\bar{q}q\rangle\delta(s)\right\}
+\frac{\langle\bar{s}s\rangle\langle\bar{s}g_s\sigma Gs\rangle}{192\pi^2}\, , \\
\rho^{22}_{f_0(980)}&=&\frac{\langle\bar{q}q\rangle^2}{144}\left\{\frac{3}{\pi^2}s+\langle\frac{\alpha_sGG}{\pi}\rangle\delta(s)+24m_s\langle\bar{s}s\rangle\delta(s)\right\}\,,\\
\rho^{22}_{a_0(980)}&=&\frac{\langle\bar{s}s\rangle^2}{288}\left\{\frac{3}{\pi^2}s+\langle\frac{\alpha_sGG}{\pi}\rangle\delta(s)+24m_q\langle\bar{q}q\rangle\delta(s)\right\}\,,
\end{eqnarray}

\begin{eqnarray}
\rho^{44}_{\kappa_0(800)}&=&\frac{s^4}{61440\pi^6}\left\{1+\frac{\alpha_s}{\pi}\left( \frac{57}{5}+2\log \frac{\mu^2}{s}\right) \right\}+\frac{(m_s-2m_q)\langle\bar{s}s\rangle-(m_q+2m_s)\langle\bar{q}q\rangle}{384\pi^4}s^2
 \nonumber\\
&&+\frac{3(m_s+m_q)\langle\bar{q}g_s\sigma Gq\rangle+(3m_q-m_s)\langle\bar{s}g_s\sigma Gs\rangle}{384\pi^4}s+\frac{\langle\bar{q}q\rangle^2+\langle\bar{q}q\rangle\langle\bar{s}s\rangle}{24\pi^2}s\nonumber\\
&&-\frac{2\langle\bar{q}q\rangle\langle\bar{q}g_s\sigma Gq\rangle+\langle\bar{q}q\rangle\langle\bar{s}g_s\sigma Gs\rangle+\langle\bar{s}s\rangle\langle\bar{q}g_s\sigma Gq\rangle}{48\pi^2}\nonumber\\
&&+\frac{\langle\bar{q}g_s\sigma Gq\rangle^2+\langle\bar{q}g_s\sigma Gq\rangle\langle\bar{s}g_s\sigma Gs\rangle}{192\pi^2}\delta(s)\nonumber\\
&&-\frac{(2m_s-m_q)\langle\bar{q}q\rangle^3+(4m_q-m_s)\langle\bar{s}s\rangle\langle\bar{q}q\rangle^2}{18}\delta(s)\nonumber\\
&&+\frac{s^2}{1536\pi^4} \langle\frac{\alpha_sGG}{\pi}\rangle+\frac{(m_s-2m_q)\langle\bar{s}s\rangle-(m_q+2m_s)\langle\bar{q}q\rangle}{384\pi^2}\langle\frac{\alpha_sGG}{\pi}\rangle\nonumber\\
&&-\frac{(2m_q+m_s)\langle\bar{q}q\rangle+m_q\langle\bar{s}s\rangle}{576\pi^2}\langle\frac{\alpha_sGG}{\pi}\rangle\nonumber\\
&&+\frac{5}{432}\left[\langle\bar{q}q\rangle^2+\langle\bar{q}q\rangle\langle\bar{s}s\rangle\right]\langle\frac{\alpha_sGG}{\pi}\rangle\delta(s)\, ,\\
\rho^{42}_{\kappa_0(800)}&=&\frac{\langle\bar{q}q\rangle^2}{288}\left\{\frac{3}{\pi^2}s+\langle\frac{\alpha_sGG}{\pi}\rangle\delta(s)
+4(m_q+2m_s)\langle\bar{q}q\rangle\delta(s)+4(m_s+2m_q)\langle\bar{s}s\rangle\delta(s)\right\}\nonumber\\
&&+\frac{\langle\bar{q}q\rangle\langle\bar{q}g_s\sigma Gq\rangle}{192\pi^2}\, , \\
\rho^{22}_{\kappa_0(800)}&=&\frac{\langle\bar{q}q\rangle^2}{288}\left\{\frac{3}{\pi^2}s+\langle\frac{\alpha_sGG}{\pi}\rangle\delta(s)
+4(m_q+2m_s)\langle\bar{q}q\rangle\delta(s)+4(m_s+2m_q)\langle\bar{s}s\rangle\delta(s)\right\}\, ,
\end{eqnarray}
\begin{eqnarray}
\rho^{44}_{f_0(500)}&=&\frac{s^4}{61440\pi^6}\left\{1+\frac{\alpha_s}{\pi}\left( \frac{57}{5}+2\log \frac{\mu^2}{s}\right) \right\}-\frac{m_q\langle\bar{q}q\rangle}{96\pi^4}s^2
+\frac{\langle\bar{q}q\rangle^2}{12\pi^2}s +\frac{m_q\langle\bar{q}g_s\sigma Gq\rangle}{48\pi^4}s\nonumber\\
&&-\frac{\langle\bar{q}q\rangle\langle\bar{q}g_s\sigma Gq\rangle}{12\pi^2}
+\frac{\langle\bar{q}g_s\sigma Gq\rangle^2}{96\pi^2}\delta(s)-\frac{2m_q\langle\bar{q}q\rangle^3}{9}\delta(s)\nonumber\\
&&+\frac{s^2}{1536\pi^4} \langle\frac{\alpha_sGG}{\pi}\rangle-\frac{5m_q\langle\bar{q}q\rangle}{288\pi^2}\langle\frac{\alpha_sGG}{\pi}\rangle
+\frac{5}{216}\langle\bar{q}q\rangle^2\langle\frac{\alpha_sGG}{\pi}\rangle\delta(s)\, ,\\
\rho^{42}_{f_0(500)}&=&\frac{\langle\bar{q}q\rangle^2}{144}\left\{\frac{3}{\pi^2}s+\langle\frac{\alpha_sGG}{\pi}\rangle\delta(s)+24m_q\langle\bar{q}q\rangle\delta(s)\right\}+\frac{\langle\bar{q}q\rangle\langle\bar{q}g_s\sigma Gq\rangle}{96\pi^2}\, , \\
\rho^{22}_{f_0(500)}&=&\frac{\langle\bar{q}q\rangle^2}{144}\left\{\frac{3}{\pi^2}s+\langle\frac{\alpha_sGG}{\pi}\rangle\delta(s)+24m_q\langle\bar{q}q\rangle\delta(s)\right\}\, .
\end{eqnarray}

We  differentiate   Eq.(21) with respect to  $-\frac{1}{M^2}$, then eliminate the
 pole residues  $\lambda_{S}$, and  obtain the QCD sum rules for
 the masses,
 \begin{eqnarray}
 m_S^2= \frac{\int_{0}^{s_S^0} ds\, \frac{d}{d(-1/M^2)}\,\rho_S(s) \,\exp\left(-\frac{s}{M^2}\right) }{\int_{0}^{s_S^0}ds \,\rho_S(s) \,\exp\left(-\frac{s}{M^2}\right)}\, .
\end{eqnarray}

\section{Numerical results and discussions}
In calculation, the input parameters are taken to be the standard values $\langle \bar{s}s
\rangle=(0.8\pm0.2)\langle \bar{q}q \rangle$, $\langle \bar{s}g_s\sigma  G
s \rangle=m_0^2\langle \bar{s}s \rangle$, $\langle \bar{q}g_s\sigma
 G q \rangle=m_0^2\langle \bar{q}q \rangle$, $m_0^2=(0.8\pm0.2)\,\rm{GeV}^2$, $\langle \bar{u}u
\rangle=\langle \bar{d}d \rangle=\langle \bar{q}q \rangle=-(0.24\pm0.01
\rm{GeV})^3$, $\langle \frac{\alpha_sGG}{\pi} \rangle=(0.33 \,\rm{GeV})^4$,
 $m_u=m_d=6\,\rm{MeV}$ and $m_s=140\,\rm{MeV}$ at the energy scale $\mu=1\,\rm{GeV}$  \cite{SVZ79,PRT85,QCDSR-review}.
  The values $m_u=m_d=6\,\rm{MeV}$ can also be obtained  from the Gell-Mann-Oakes-Renner relation at the energy scale $\mu=1\,\rm{GeV}$ in the isospin limit.

 \begin{table}
\begin{center}
\begin{tabular}{|c|c|c|c|c|}\hline\hline
                  & $m_{S}\,(\rm{MeV})$   & $\Gamma_S\,( \rm{MeV})$    & $m_{S}+\Gamma_S/2\,( \rm{MeV})$    & $m_{S}-\Gamma_S/2\,( \rm{MeV})$   \\ \hline
   $f_0(980)$     & $990 \pm20$           & $40-100$                   & $1025^c$                           &                          \\ \hline
 $f_0(1500)$      & $1504\pm 6$           & $109\pm 7$                 &                                    & $1450^c$              \\ \hline
   $a_0(980)$     & $980\pm20$            & $50-100$                   & $1018^c$                           &                       \\ \hline
  $a_0(1450)$     & $1474 \pm19$          & $265 \pm13$                &                                    & $1342^c$              \\ \hline
$\kappa_0(800)$   & $682 \pm29$           & $547\pm24$                 & $956^c$                            &                      \\ \hline
 $K^*_0(1430)$    & $1425\pm50$           & $270 \pm80$                &                                    & $1290^c$             \\ \hline
   $f_0(500)$     & $400-550$             & $400-700$                  & $750^c$                            &                      \\ \hline
 $f_0(1370)$      & $1200-1500$           & $200-500$                  &                                    & $1250^*$            \\ \hline
 \hline
\end{tabular}
\end{center}
\caption{ The Breit-Wigner  masses and widths  of the scalar  mesons from the Particle Data Group, where the superscript  $c$ denotes  the central values, and the superscript  $*$ denotes that we have taken the low bound of the width of the $f_0(1370)$.  }
\end{table}

 Firstly, let us set the mixing angles $\theta_S$ in the QCD spectral densities $\rho_S(s)$ in Eq.(22) to be zero, then the scalar nonet mesons are pure tetraquark states. The perturbative QCD spectral densities are proportional to $s^4$, it is difficult  to satisfy the pole dominance condition ${\rm PC}\geq 50\%$  if the continuum threshold parameters $s_S^0$  are not large  enough and the Borel parameters $M^2$ are not small enough, where the pole contribution (PC) is defined by
 \begin{eqnarray}
 {\rm PC} &=& \frac{\int_0^{s^0_S} ds \, \rho_S(s)\, \exp\left( -\frac{s}{M^2}\right)}{\int_0^{\infty} ds \, \rho_S(s)\, \exp\left( -\frac{s}{M^2}\right)}\, .
 \end{eqnarray}

   For $s_S^0$, it is reasonable to take any values satisfying the relation, $m_{\rm gr}+\frac{\Gamma_{\rm gr}}{2}\leq \sqrt{s^0_S}\leq m_{\rm 1st}-\frac{\Gamma_{\rm 1st}}{2}$, where the gr and 1st denote the ground state and the first excited state (or the higher resonant state) respectively. The $\sqrt{s^0_S}$ \, lies between the two Breit-Wigner resonances,  if we parameterize  the scalar mesons with the  Breit-Wigner masses and widths. More explicitly,
  \begin{eqnarray}
  m_{f_0(980)}+\frac{\Gamma_{f_0(980)}}{2} \leq &\sqrt{s^0_{f_{0}(980)}}&\leq m_{f_0(1500)}-\frac{\Gamma_{f_0(1500)}}{2} \, , \nonumber\\
  m_{a_0(980)}+\frac{\Gamma_{a_0(980)}}{2} \leq &\sqrt{s^0_{a_{0}(980)}}&\leq m_{a_0(1450)}-\frac{\Gamma_{a_0(1450)}}{2} \, , \nonumber\\
  m_{\kappa_0(800)}+\frac{\Gamma_{\kappa_0(800)}}{2} \leq &\sqrt{s^0_{\kappa_{0}(800)}}&\leq m_{K^*_0(1430)}-\frac{\Gamma_{K^*_0(1430)}}{2} \, , \nonumber\\
   m_{f_0(500)}+\frac{\Gamma_{f_0(500)}}{2} \leq &\sqrt{s^0_{f_{0}(500)}}&\leq m_{f_0(1370)}-\frac{\Gamma_{f_0(1370)}}{2} \, .
  \end{eqnarray}
In Table 1, we show the Breit-Wigner  masses and widths  of the scalar  mesons from the Particle Data Group explicitly \cite{PDG}.
 Based on the values in Table 1, we can choose the largest  continuum threshold parameters
  $s^0_{f_0(980)}=1.9\,\rm{GeV}^2$,  $s^0_{a_0(980)}=1.8\,\rm{GeV}^2$,   $s^0_{\kappa_0(800)}=1.7\,\rm{GeV}^2$ and   $s^0_{f_0(500)}=1.6\,\rm{GeV}^2$ tentatively to take into account all the ground state contributions and  avoid the possible contaminations from the higher resonances $f_0(1370)$, $a_0(1450)$, $K^*_0(1430)$ and $f_0(1500)$.

  In Fig.3, we plot the masses of the scalar mesons as pure tetraquark states  with variations of the  Borel  parameter $M^2$, where the central values of other parameters are taken. From the figure, we can see that if we exclude the contributions of the condensates $\langle \bar{q}q\rangle\langle \bar{q}^\prime g_s\sigma Gq^\prime\rangle$ with $q,q^\prime=u,d,s$, the predicted masses $m_S$ increase monotonously and quickly with increase of the Borel parameters  $M^2$ at the value $M^2< 0.9\,\rm{GeV}^2$, then increase  slowly and reach the  values $m_{f_0(980)}=1.06\,\rm{GeV}$, $m_{a_0(980)}=1.03\,\rm{GeV}$, $m_{\kappa_0(800)}=0.99\,\rm{GeV}$, $m_{f_0(500)}=0.96\,\rm{GeV}$ at the value  $M^2=3.3\,\rm{GeV}^2$. It is possible to reproduce the experimental data with fine tuning the continuum threshold parameters.  However, if we include the contributions of the  condensates $\langle \bar{q}q\rangle\langle \bar{q}^\prime g_s\sigma Gq^\prime\rangle$, the predicted masses $m_S$ are amplified greatly. The $m_S$ decrease monotonously and quickly  with increase of the Borel parameters $M^2$  below some special values, for example, $M^2< 1.2\,\rm{GeV}^2$ for the $f_0(980)$ and $a_0(980)$, then decrease  slowly and reach the
 values  $m_S\geq 1.4\,\rm{GeV}$ at the value  $M^2=3.3\,\rm{GeV}^2$. It is impossible to reproduce the experimental data by  fine tuning the   continuum threshold parameters.
In Fig.4, we plot the contributions of different terms
    in the operator product expansion  with variations of the
  Borel parameters $M^2$ for the scalar nonet mesons as the pure tetraquark states. From the figure, we can see that the convergent behavior of the operator product expansion is very bad, for example,  the condensates $\langle \bar{q}q\rangle\langle \bar{q}^\prime g_s\sigma Gq^\prime\rangle$  of dimension 8 with $q,q^\prime=u,d,s$ have too large negative values  at the region $M^2\geq 1.2\,\rm{GeV}^2$.
From Figs.3-4, we can draw the conclusion tentatively that the condensates$\langle \bar{q}q\rangle\langle \bar{q}^\prime g_s\sigma Gq^\prime\rangle$  of dimension 8 play an important role. The conclusion  is compatible with the observation of Ref.\cite{Lee2006}, that there exists   no evidence of the couplings of the tetraquark states to the pure light scalar   nonet mesons \cite{Lee2006}.

\begin{figure}
 \centering
  \includegraphics[totalheight=5cm,width=7cm]{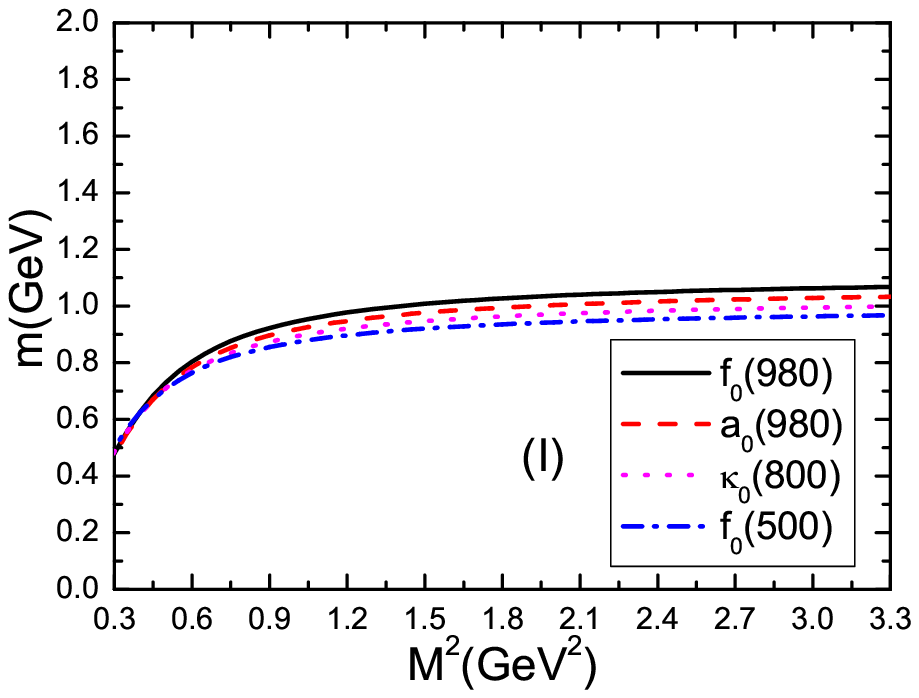}
  \includegraphics[totalheight=5cm,width=7cm]{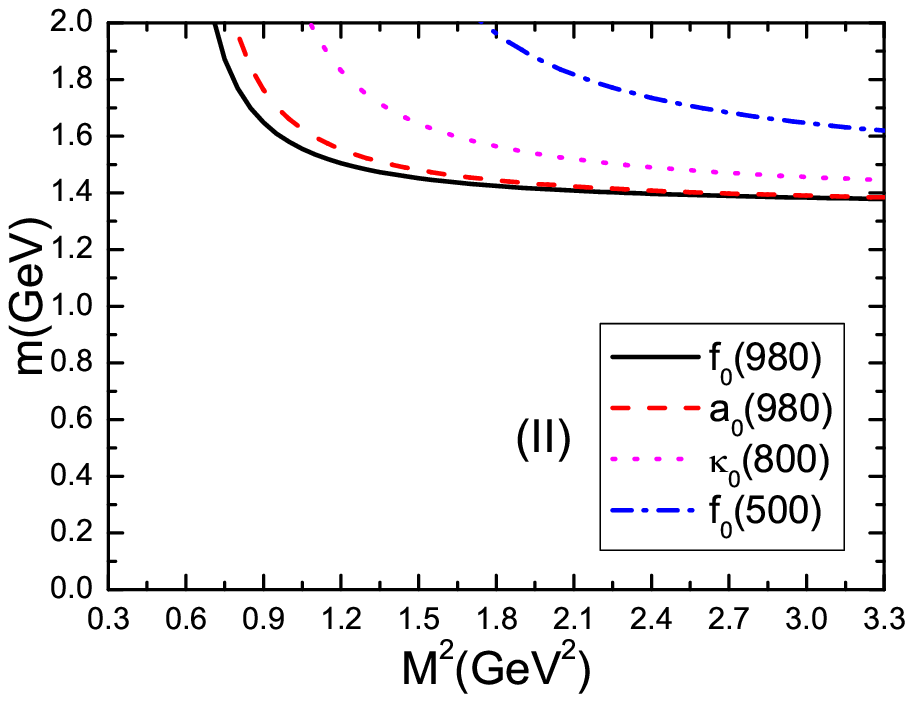}
     \caption{The masses of the scalar mesons as pure tetraquark states  with variations of the  Borel  parameter $M^2$, where the (I) and (II) denote the contributions of the condensates $\langle \bar{q}q\rangle\langle \bar{q}^\prime g_s\sigma Gq^\prime\rangle$ of dimension 8 are excluded and included, respectively, $q,q^\prime=u,d,s$.}
\end{figure}

\begin{figure}
 \centering
  \includegraphics[totalheight=5cm,width=7cm]{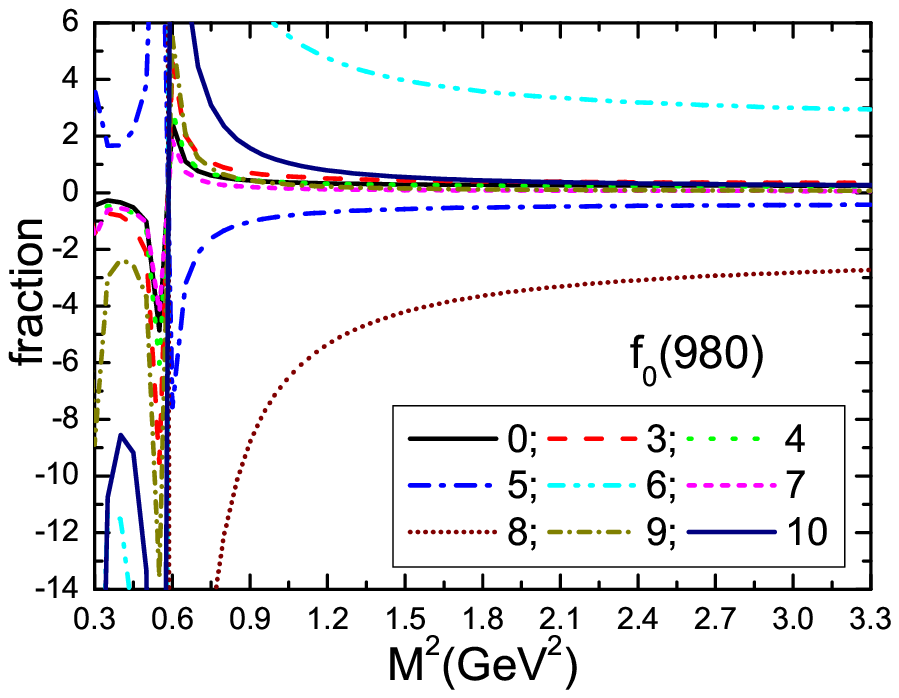}
 \includegraphics[totalheight=5cm,width=7cm]{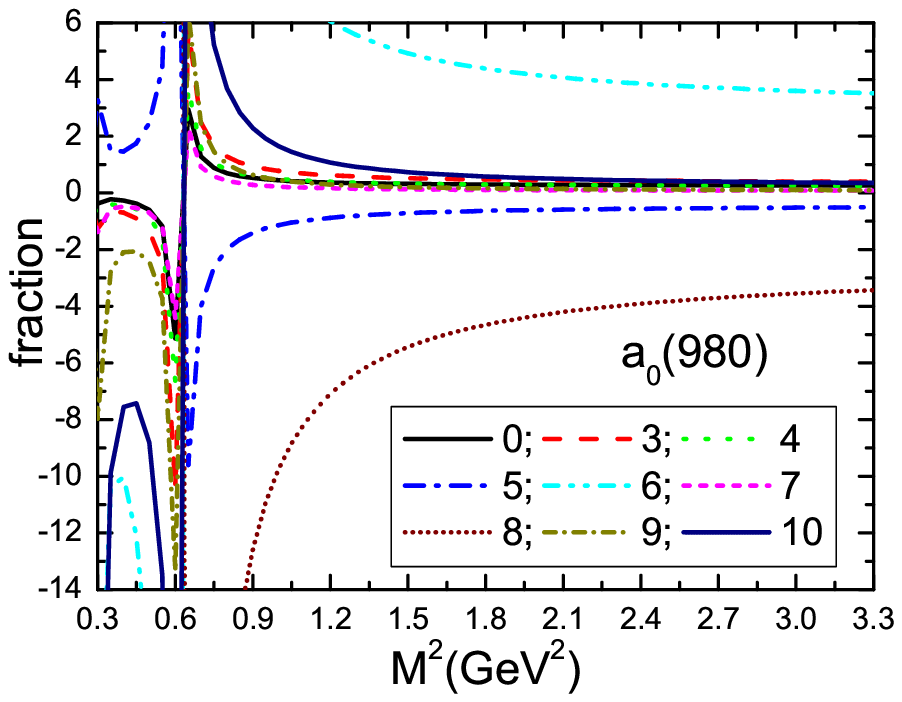}
 \includegraphics[totalheight=5cm,width=7cm]{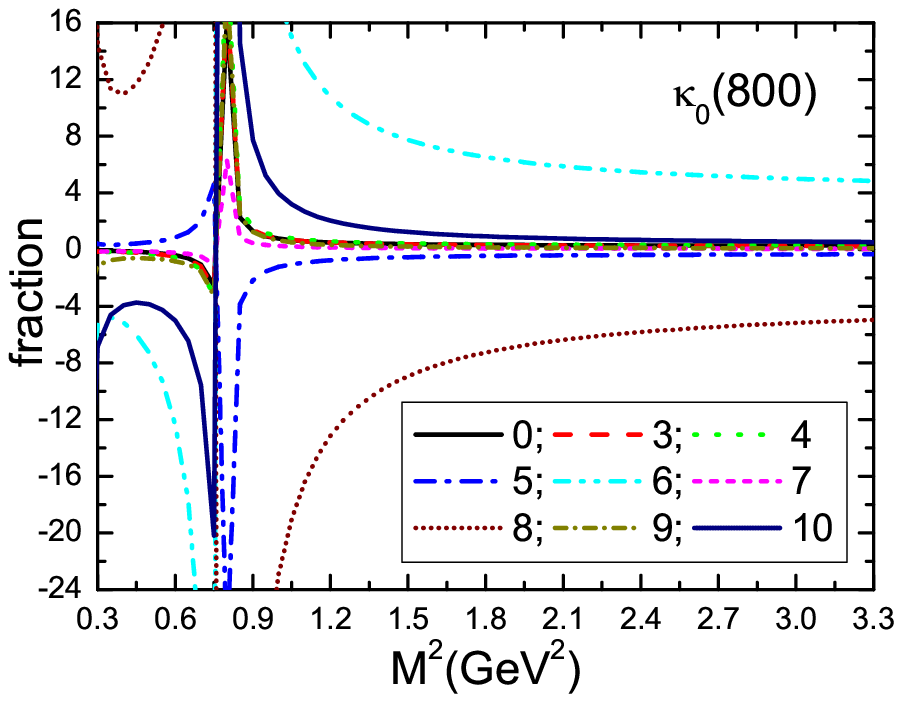}
 \includegraphics[totalheight=5cm,width=7cm]{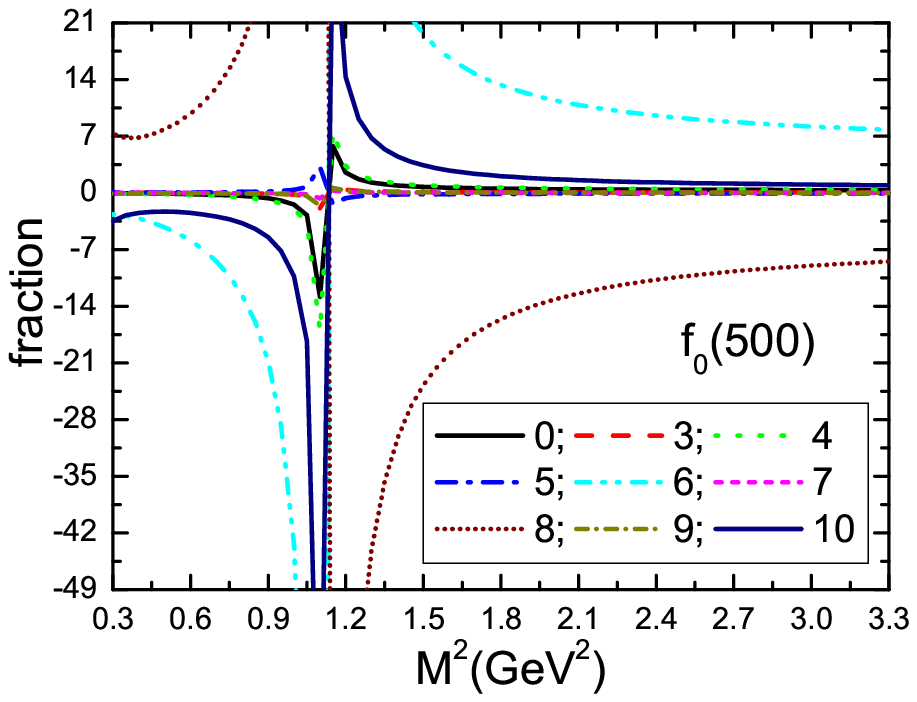}
    \caption{The contributions of different terms
    in the operator product expansion  with variations of the
  Borel parameter $M^2$ for the scalar nonet mesons as pure tetraquark states, where the $0$, $3$, $4$, $5$, $6$, $7$, $8$, $9$ and $10$   denotes the dimensions of the vacuum condensates.}
\end{figure}

Now we set the mixing angles  $\theta_S$ to be $90^{\circ}$ in the QCD spectral densities $\rho_S(s)$ in Eq.(22), and take the scalar nonet mesons to be  pure two-quark states.  In Fig.5, we plot the masses of the scalar mesons as pure two-quark states  with variations of the  Borel  parameters  $M^2$, the same parameters as that in Fig.3 are taken. From the figure, we can see that the predicted masses $m_S\approx (0.85-1.14)\,\rm{GeV}$ at the value $M^2=(0.5-3.3)\,\rm{GeV}^2$, there also exist some difficulty   to reproduce the  experimental data approximately by fine tuning the continuum threshold parameters. In Fig.6, we plot the contributions of different terms
    in the operator product expansion  with variations of the
  Borel parameters  $M^2$ for the scalar nonet mesons as the pure two-quark states. From the figure, we can see that the convergent behavior of the operator product expansion is very good, the main contributions come from the perturbative terms, which are of dimension 6 according to the normalization factors $\langle \bar{q}q\rangle^2$ and $\langle \bar{s}s\rangle^2$.

\begin{figure}
 \centering
    \includegraphics[totalheight=6cm,width=8cm]{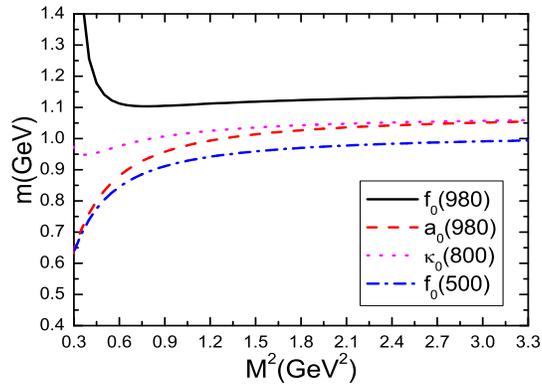}
    \caption{The masses of the scalar mesons as pure two-quark states  with variations of the  Borel  parameter $M^2$.}
\end{figure}

\begin{figure}
 \centering
  \includegraphics[totalheight=5cm,width=7cm]{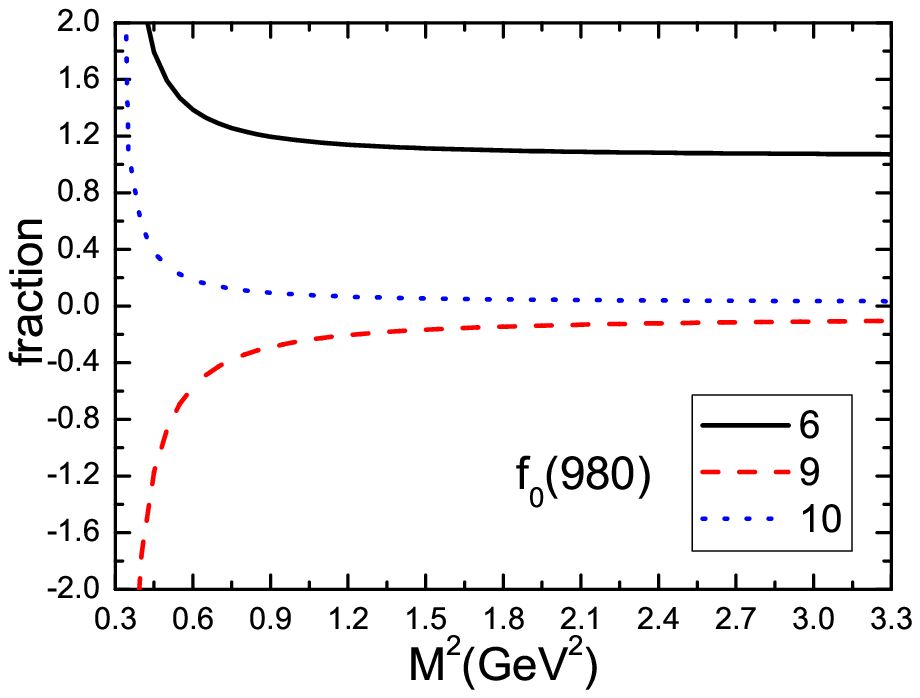}
 \includegraphics[totalheight=5cm,width=7cm]{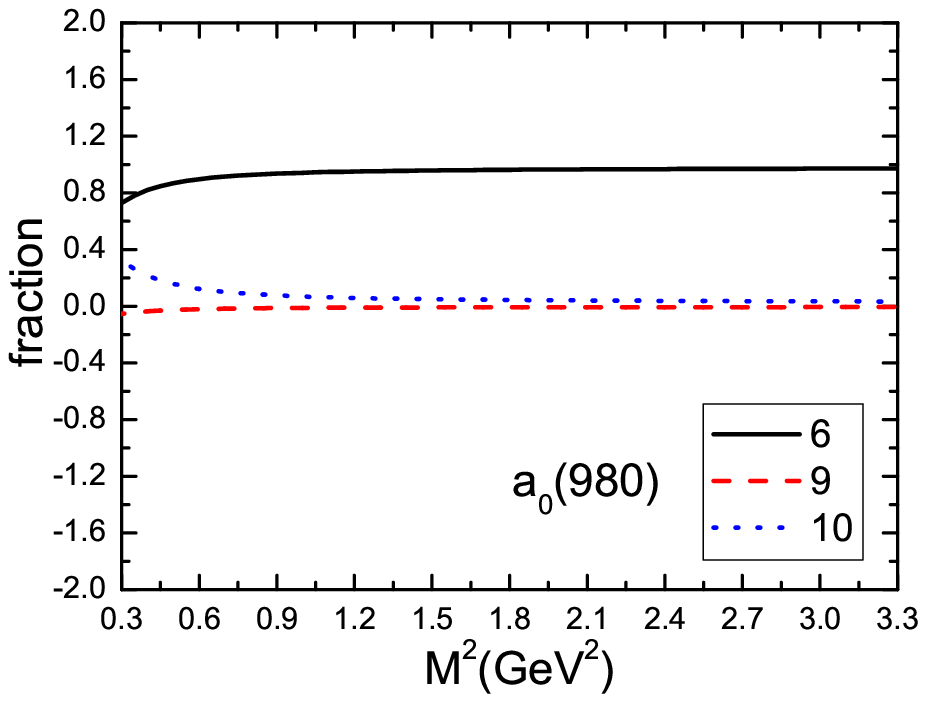}
 \includegraphics[totalheight=5cm,width=7cm]{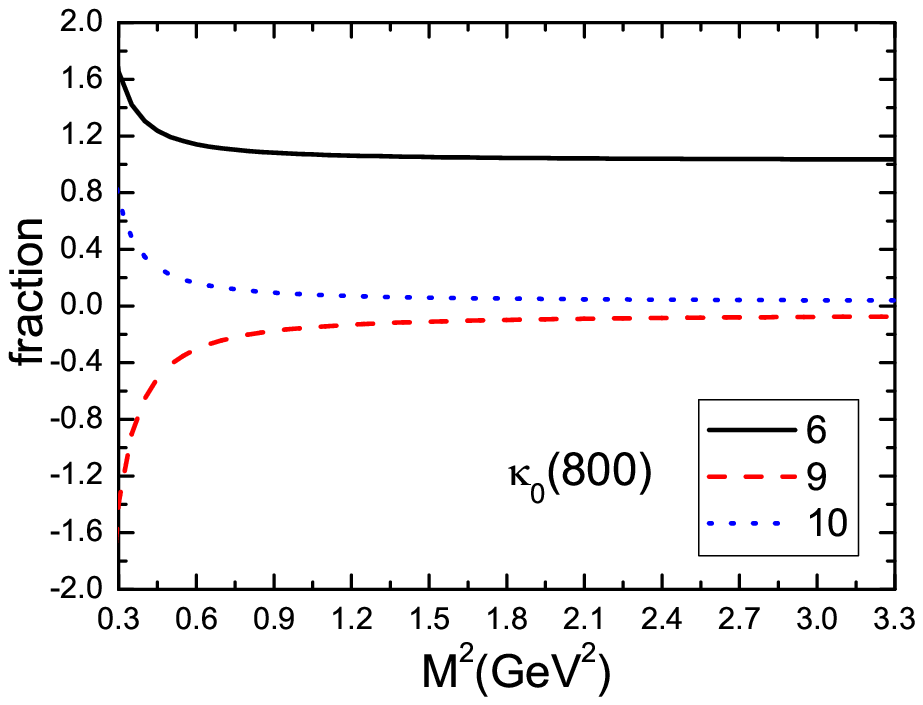}
 \includegraphics[totalheight=5cm,width=7cm]{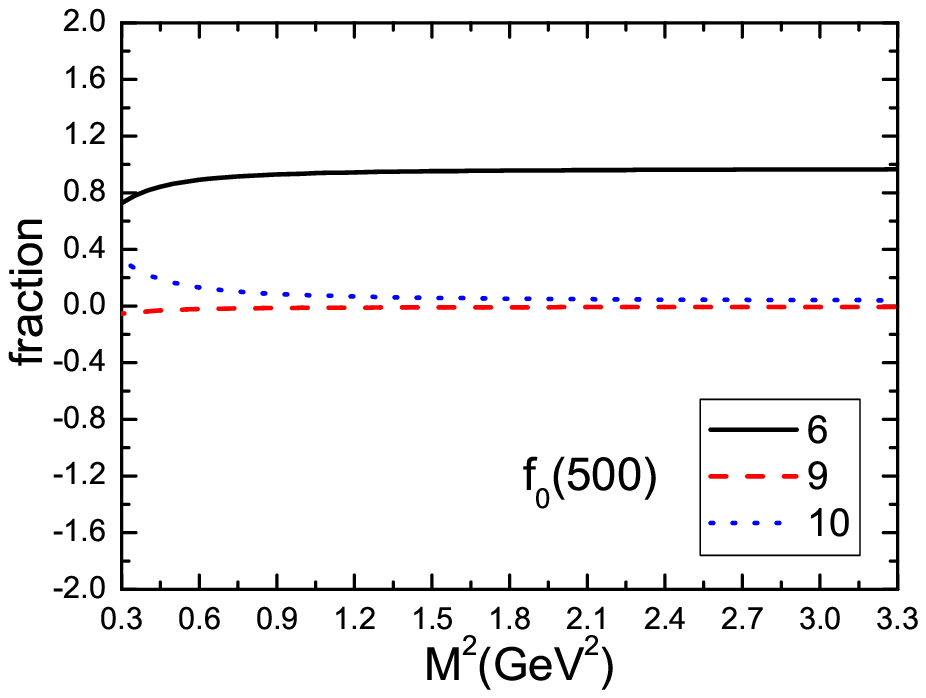}
    \caption{The contributions of different terms
    in the operator product expansion  with variations of the
  Borel parameter $M^2$ for the scalar nonet mesons as  pure two-quark states, where the  $6$, $9$ and $10$   denotes the dimensions of the vacuum condensates. We have taken into account the normalization factors $\langle \bar{q}q\rangle^2$ and $\langle \bar{s}s\rangle^2$.}
\end{figure}

We turn on the mixing angles  $\theta_S\neq 0^{\circ},\,90^{\circ}$ and take into account all the Feynman diagrams  which contribute to the condensate
$\langle \bar{q}q\rangle\langle \bar{q}^\prime g_s\sigma Gq^\prime\rangle$ with $q,q^\prime=u,d,s$, see the Feynman diagrams in Figs.1-2.   The contributions of the vacuum condensates $\langle \bar{q}q\rangle\langle \bar{q}^\prime g_s\sigma Gq^\prime\rangle$ of dimension 8 can be canceled out completely with the ideal mixing angles $\theta^0_S$,
 \begin{eqnarray}
 \theta^0_{f_0(980)}&=& \tan^{-1}\left(2\frac{\langle\bar{q}q\rangle\langle\bar{s}g_s\sigma Gs\rangle+\langle\bar{s}s\rangle\langle\bar{q}g_s\sigma Gq\rangle}{\langle\bar{q}q\rangle\langle\bar{q}g_s\sigma Gq\rangle}\right) \approx 72.6^{\circ}\, ,\nonumber\\
  \theta^0_{a_0(980)}&=&\tan^{-1}\left(4\frac{\langle\bar{q}q\rangle\langle\bar{s}g_s\sigma Gs\rangle+\langle\bar{s}s\rangle\langle\bar{q}g_s\sigma Gq\rangle}{\langle\bar{s}s\rangle\langle\bar{s}g_s\sigma Gs\rangle}\right) \approx 84.3^{\circ}\, ,\nonumber\\
  \theta^0_{\kappa_0(800)}&=&\tan^{-1}\left(2\frac{2\langle\bar{q}q\rangle\langle\bar{q}g_s\sigma Gq\rangle+\langle\bar{q}q\rangle\langle\bar{s}g_s\sigma Gs\rangle+\langle\bar{s}s\rangle\langle\bar{q}g_s\sigma Gq\rangle}{\langle\bar{q}q\rangle\langle\bar{q}g_s\sigma Gq\rangle} \right)\approx 82.1^{\circ}\, ,\nonumber\\
  \theta^0_{f_0(500)}&=&\tan^{-1}\left(4\right) \approx 76.0^{\circ}\, ,
 \end{eqnarray}
which results  in much better convergent  behavior in the operator product expansion.

In this article, we choose the mixing angles $\theta_S=\theta_S^0$, then  impose
the two criteria (i.e. pole dominance and convergence of the operator product
expansion) of the QCD sum rules on the two-quark-tetraquark mixed states, and search for the optimal  values of the Borel parameters $M^2$ and continuum threshold
parameters $s^0_S$.
The resulting Borel parameters (or Borel windows), continuum threshold parameters and pole contributions of the scalar nonet mesons are shown in Table 2 explicitly.

From Table 2, we can see that the upper bound of the pole contributions can reach $(51-69)\%$, the pole dominance condition is satisfied marginally. If we intend to obtain QCD sum rules for the light tetraquark states with the pole contributions larger than $50\%$, we  should resort to ¡±multi-pole
plus  continuum states¡± to approximate the phenomenological spectral densities, include at least the  ground state plus the first excited state, and postpone the continuum threshold parameters $s_S^0$ to much larger values  \cite{WangNPA}.  In this article, we exclude  the contaminations of the continuum states  by the truncation $s_S^0$, see Eq.(34), although the truncation $s_S^0$ cannot lead to the pole contribution larger than (or about) $50\%$ in all the Borel windows.
 Such a situation is in contrary to the hidden-charm and hidden-bottom tetraquark states and hidden-charm pentaquark states, where  the two heavy quarks $Q$ and $\bar{Q}$ stabilize the four-quark systems $q\bar{q}^{\prime}Q\bar{Q}$ and five-quark systems $qq^{\prime}q^{\prime\prime}Q\bar{Q}$, and result in QCD sum rules satisfying the pole dominance condition \cite{WangIJMPLA,Wang-QQ-tetraquark,Wang-QQ-pentaquark}.

\begin{figure}
 \centering
  \includegraphics[totalheight=5cm,width=7cm]{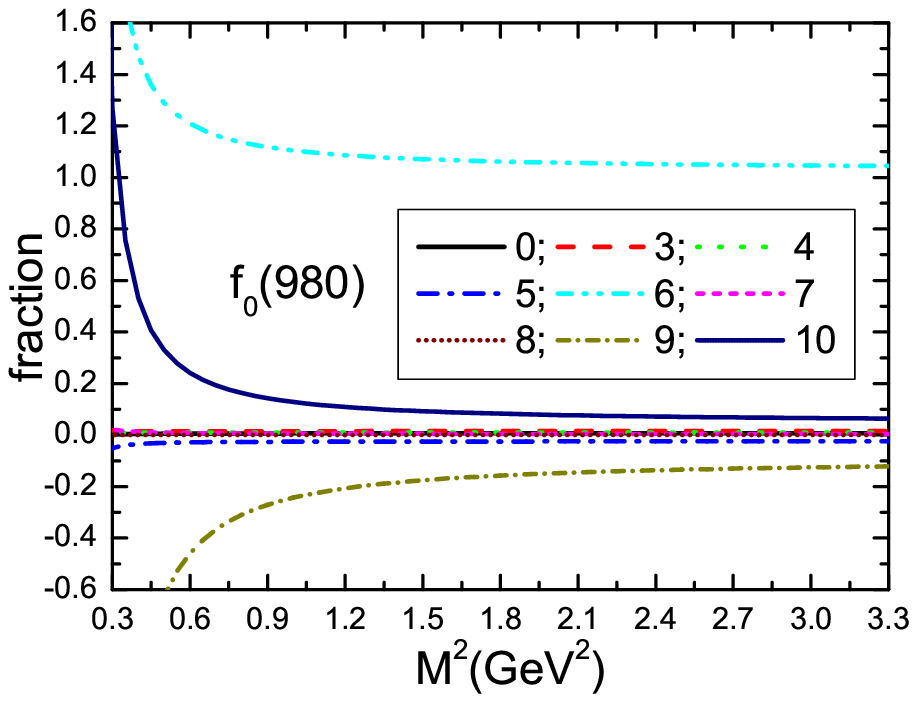}
 \includegraphics[totalheight=5cm,width=7cm]{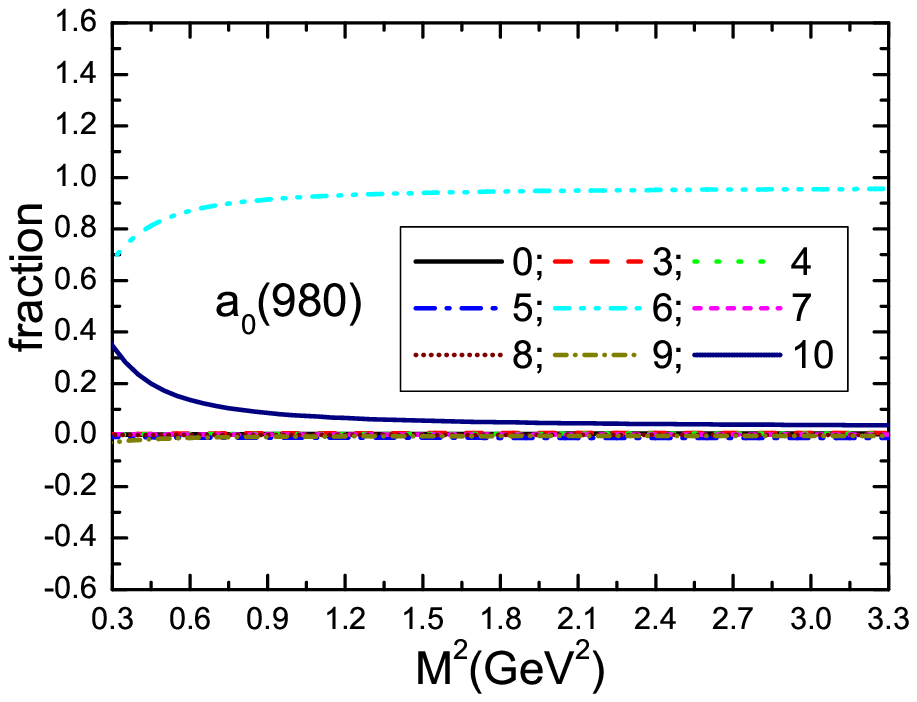}
 \includegraphics[totalheight=5cm,width=7cm]{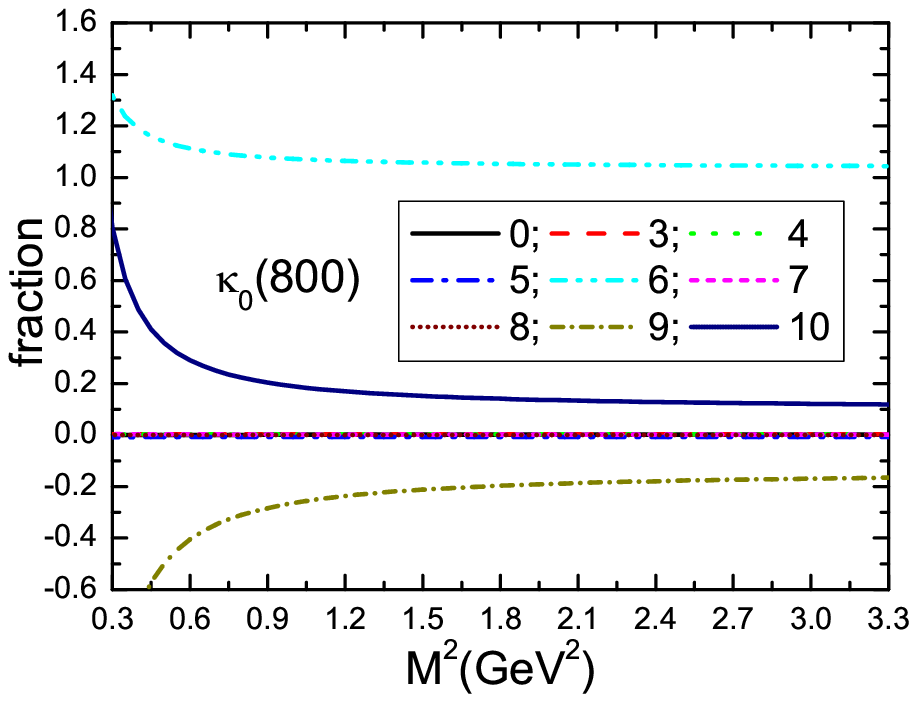}
 \includegraphics[totalheight=5cm,width=7cm]{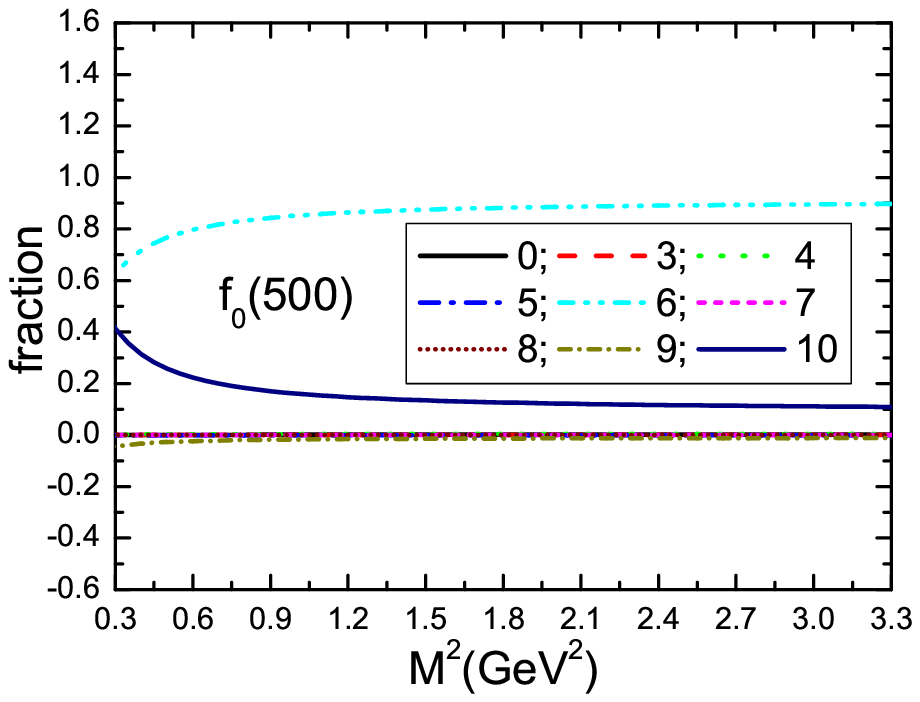}
    \caption{The contributions of different terms
    in the operator product expansion  with variations of the
  Borel parameter $M^2$ for the scalar nonet mesons as  two-quark-tetraquark mixed states, where the $0$, $3$, $4$, $5$, $6$, $7$, $8$, $9$ and $10$   denotes the dimensions of the vacuum condensates.}
\end{figure}

In Fig.7, we plot the contributions of different terms
    in the operator product expansion  with variations of the
  Borel parameter $M^2$ for the scalar nonet mesons as the two-quark-tetraquark mixed states, where the central values of other parameters are taken. From the figure, we can see that the dominant  contributions come from the vacuum condensates of dimension 6. The perturbative contributions of the two-quark components $\Pi^{22}_S(p)$ of the correlation functions $\Pi_S(p)$ are proportional to the vacuum condensate $\langle\bar{q}q\rangle^2$ (or $\langle\bar{s}s\rangle^2$) of dimension 6 according to the normalization factors $\langle \bar{q}q\rangle$ (or $\langle \bar{s}s\rangle$) in the interpolating currents $J^2_S(x)$. In the Borel windows, the contributions of the vacuum condensates of dimension 6 are about $(109-114)\%$,   $(90-93)\%$,  $(107-111)\%$ and $(80-85)\%$ for the $f_0(980)$, $a_0(980)$, $\kappa_0(800)$ and $f_0(500)$, respectively;
  the contributions of the vacuum condensates of dimension 10 are about $(11-16)\%$, $(7-10)\%$, $(19-29)\%$ and $(16-22)\%$ for the $f_0(980)$, $a_0(980)$, $\kappa_0(800)$ and $f_0(500)$, respectively, where the total contributions are normalized to be $1$. The operator product expansion is well convergent in the Borel windows shown in Table 2.

\begin{figure}
 \centering
  \includegraphics[totalheight=5cm,width=7cm]{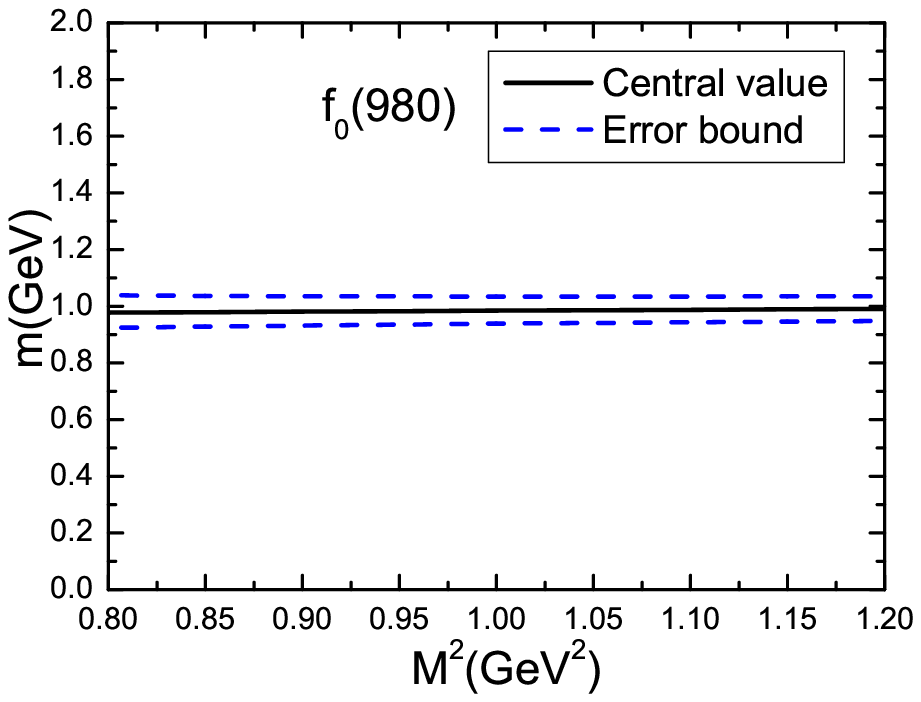}
 \includegraphics[totalheight=5cm,width=7cm]{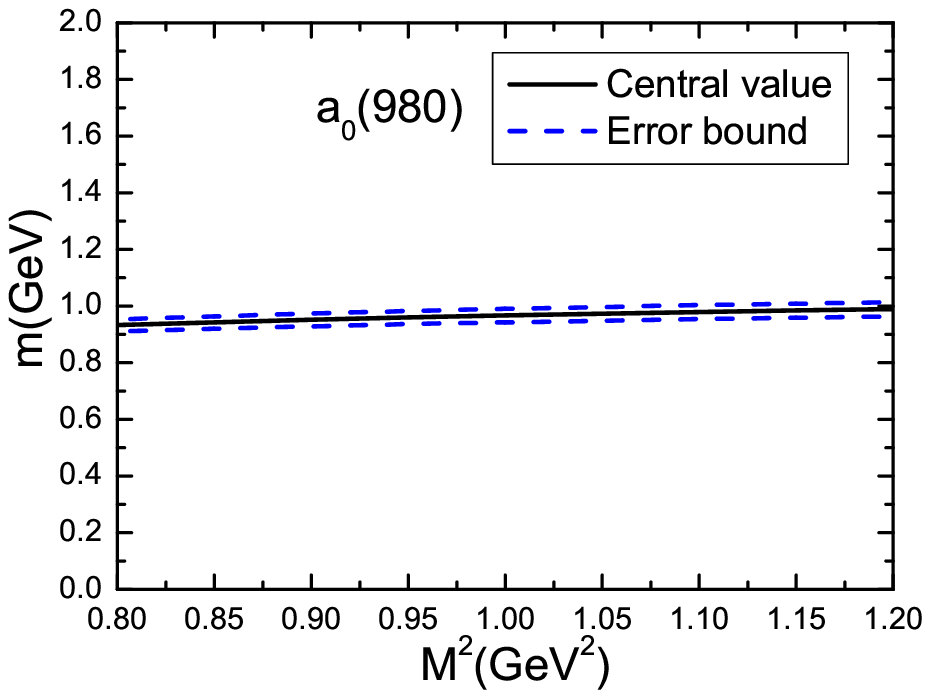}
 \includegraphics[totalheight=5cm,width=7cm]{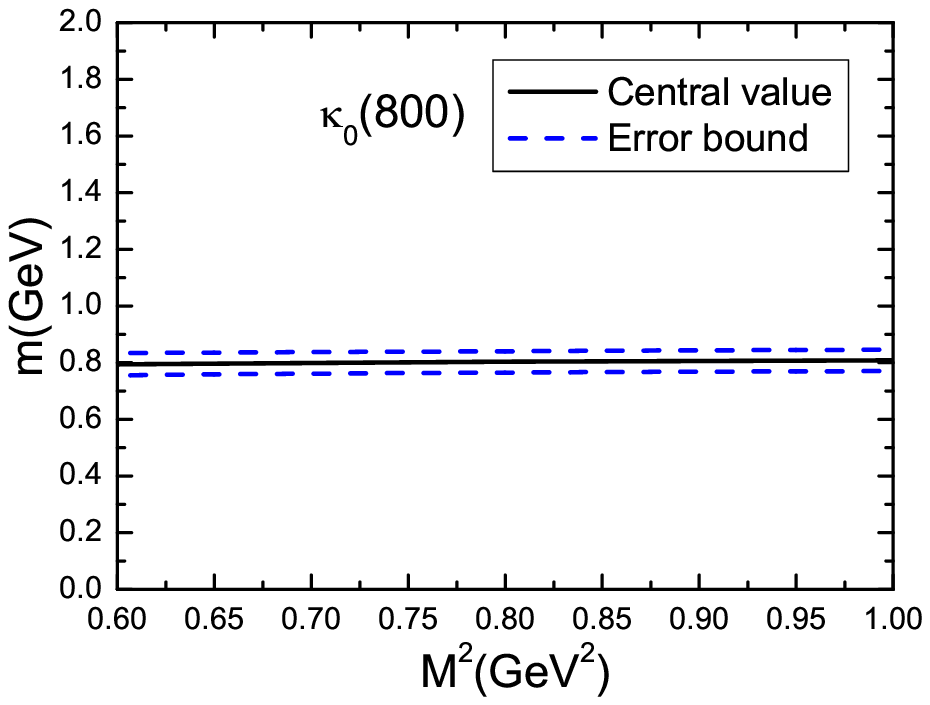}
 \includegraphics[totalheight=5cm,width=7cm]{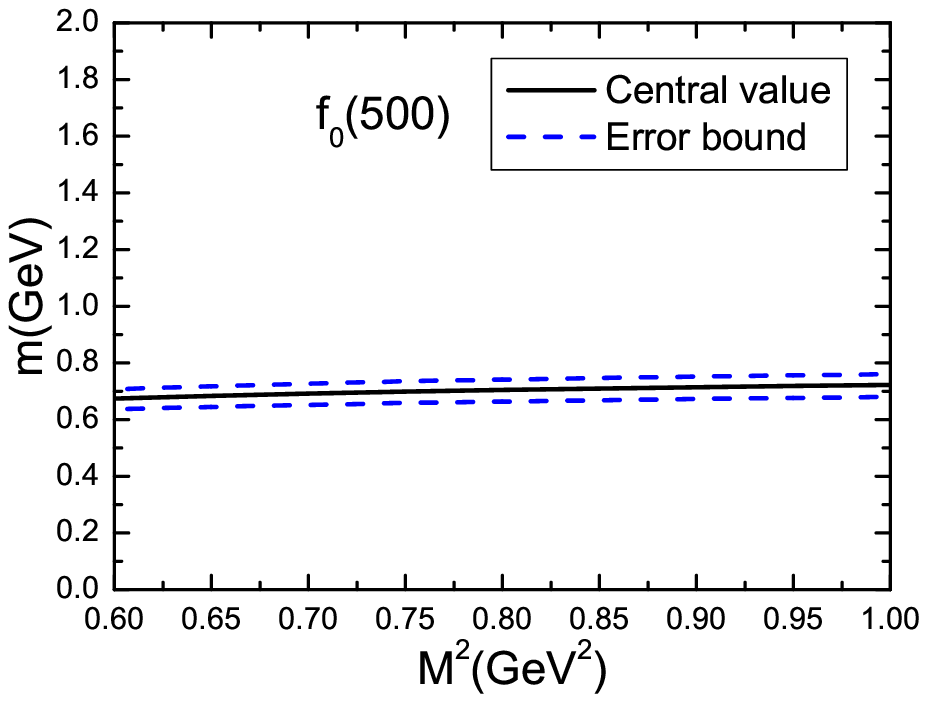}
    \caption{The masses of the scalar nonet mesons as the two-quark-tetraquark mixed states  with variations of the Borel parameters.}
\end{figure}

\begin{figure}
 \centering
  \includegraphics[totalheight=5cm,width=7cm]{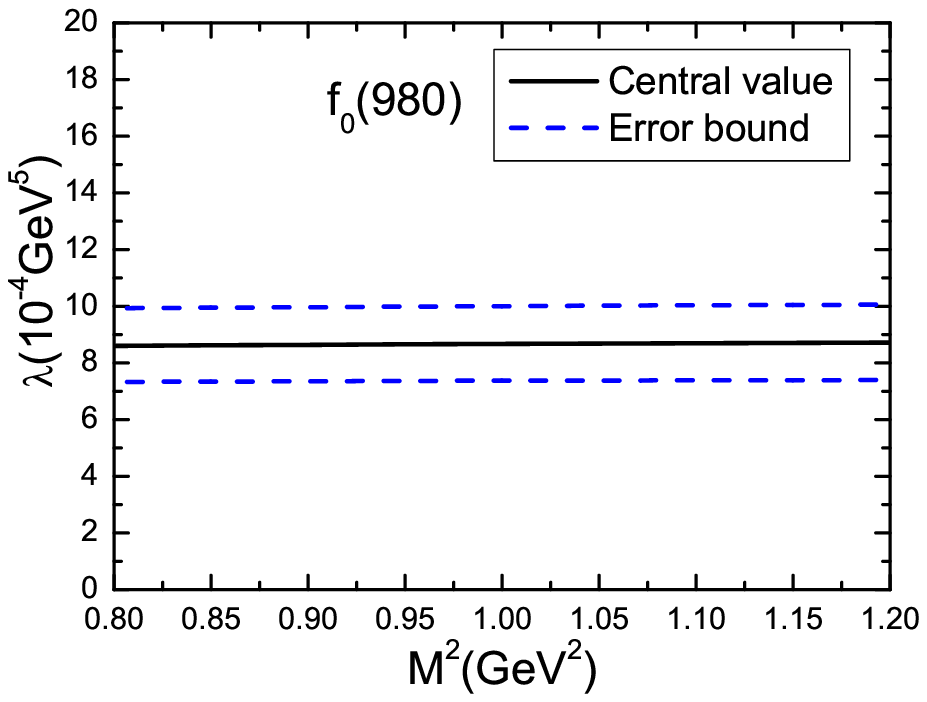}
  \includegraphics[totalheight=5cm,width=7cm]{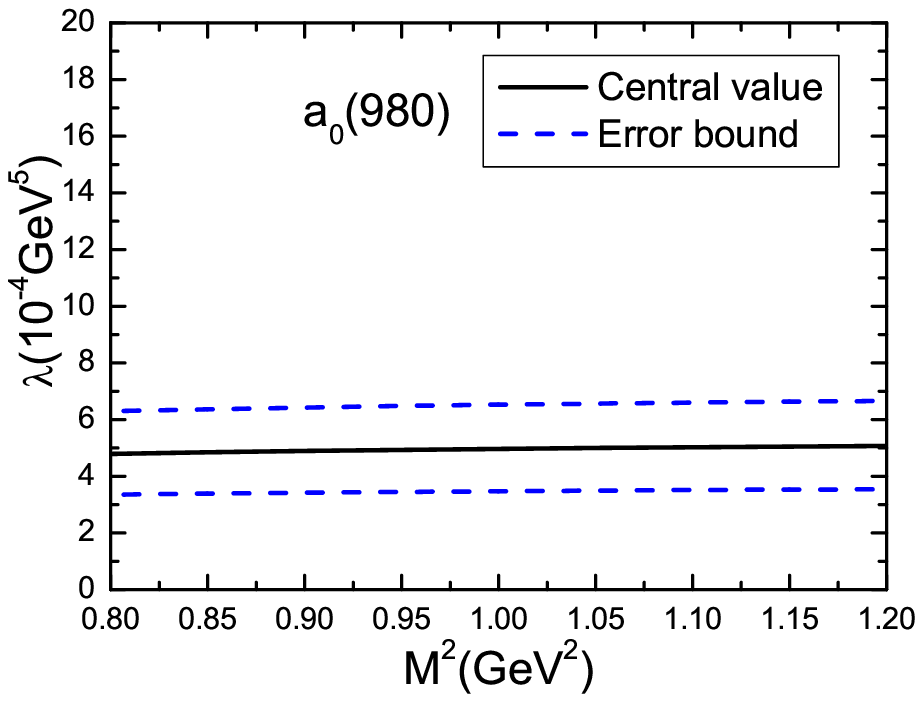}
  \includegraphics[totalheight=5cm,width=7cm]{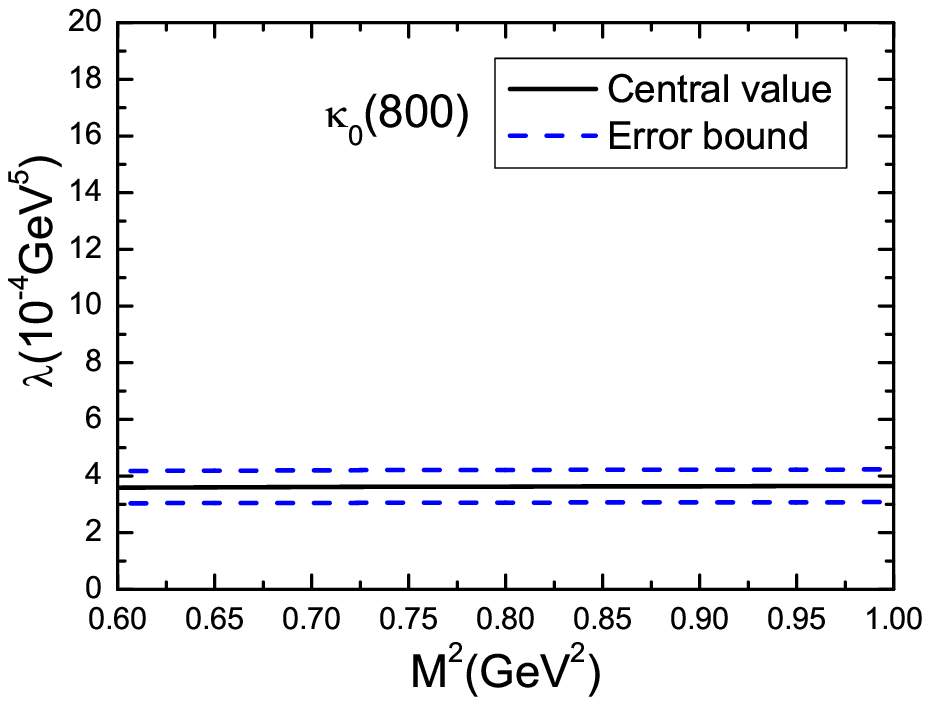}
  \includegraphics[totalheight=5cm,width=7cm]{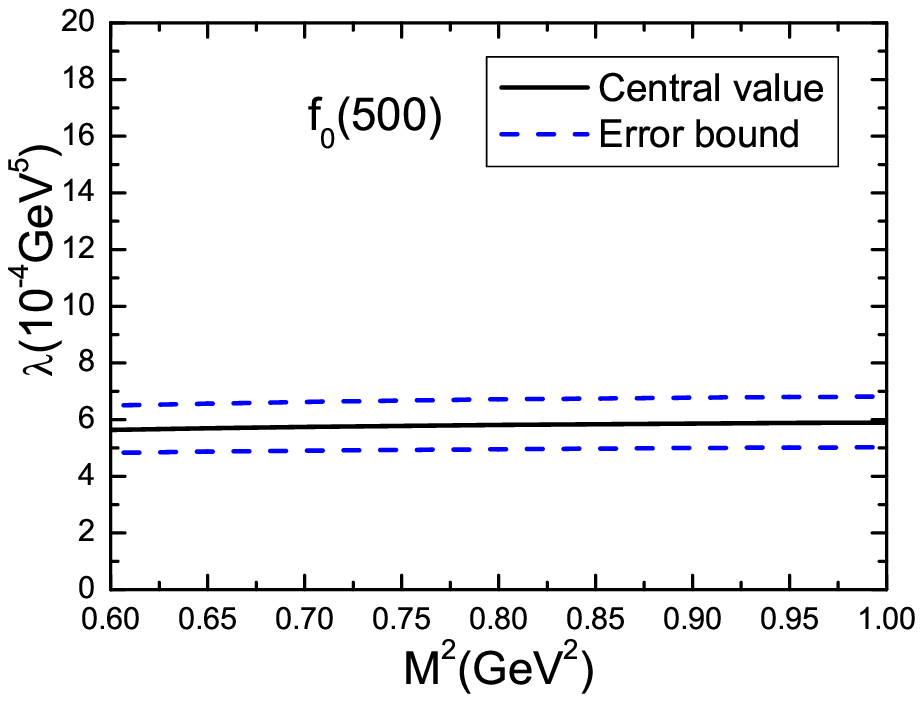}
    \caption{The pole residues   of the scalar nonet mesons as the two-quark-tetraquark mixed states  with variations of the Borel parameters.}
\end{figure}

Now we can see that it is reasonable to extract the masses from the QCD sum rules by choosing the Borel parameters and continuum threshold parameters  shown in Table 2.
In Figs.8-9, we plot the masses and pole residues of the scalar nonet mesons as the two-quark-tetraquark mixed states with variations of the Borel parameters in the Borel windows by taking into account the uncertainties of the input parameters. From the figures, we can see that the platforms are very flat, the predictions are reliable. In Table 3, we present the masses and pole residues of the scalar nonet mesons as the two-quark-tetraquark mixed states, where all uncertainties of the input parameters are taken into account.

\begin{figure}
 \centering
    \includegraphics[totalheight=6cm,width=8cm]{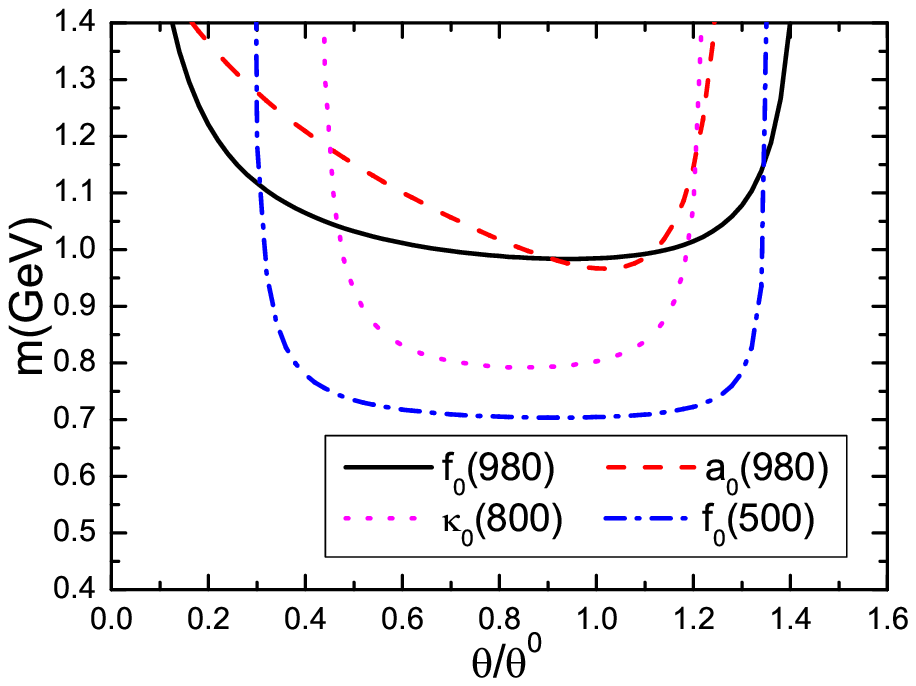}
    \caption{The masses of the scalar mesons as  two-quark-tetraquark mixed states  with variations of the  mixing  angle $\theta_S$.}
\end{figure}
\begin{figure}
 \centering
    \includegraphics[totalheight=6cm,width=8cm]{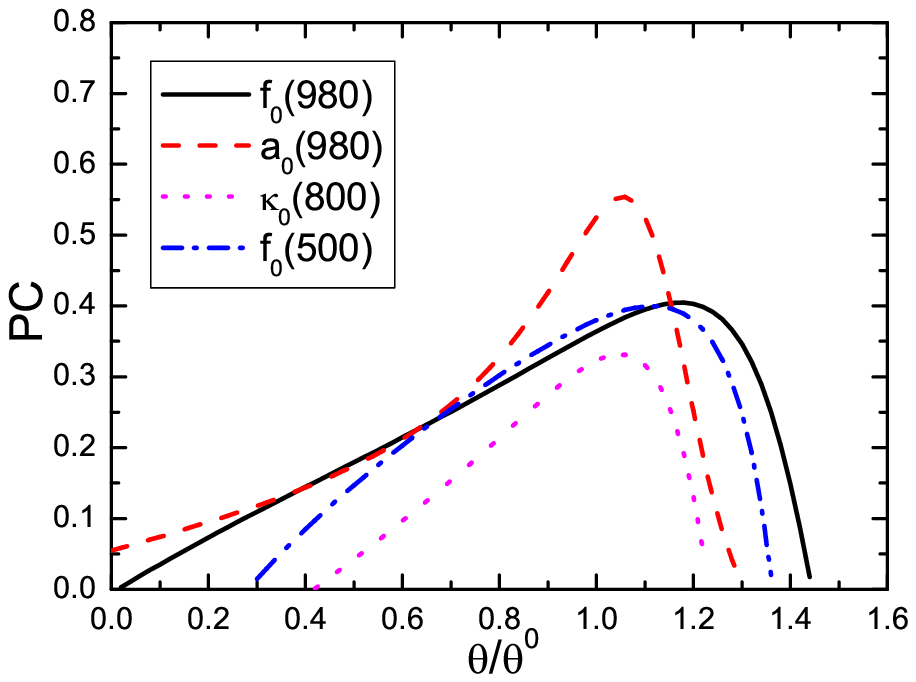}
    \caption{The pole contributions (PC) of the scalar mesons as  two-quark-tetraquark mixed states  with variations of the  mixing  angle $\theta_S$.}
\end{figure}

There exists  a compromise between the minimal masses  and the maximal pole contributions, and in the following two paragraphs  we will show that the mixing  angles $\theta_S^0$ are  optimal values.

In Fig.10, we plot the masses of the scalar mesons as the  two-quark-tetraquark mixed states  with variations of the  mixing  angles $\theta_S$,
where the input parameters are chosen as $s^0_{f_0(980)}=1.5\,\rm{GeV}^2$, $M^2_{f_0(980)}=1\,\rm{GeV}^2$,  $s^0_{a_0(980)}=1.8\,\rm{GeV}^2$, $M^2_{a_0(980)}=1\,\rm{GeV}^2$,  $s^0_{\kappa_0(800)}=1.0\,\rm{GeV}^2$, $M^2_{\kappa_0(800)}=0.8\,\rm{GeV}^2$,    $s^0_{f_0(500)}=1.0\,\rm{GeV}^2$, $M^2_{f_0(500)}=0.8\,\rm{GeV}^2$, we introduce the subscripts $f_0(980)$, $a_0(980)$, $\kappa_0(800)$ and $f_0(500)$ to denote the different  Borel parameters. From the figure, we can see that there appear minima in the predicted masses at the values $\theta_{f_0(980)}/\theta_{f_0(980)}^0=0.6-1.2$,
 $\theta_{a_0(980)}/\theta_{a_0(980)}^0=0.9-1.1$, $\theta_{\kappa_0(800)}/\theta_{\kappa_0(800)}^0=0.6-1.1$, $\theta_{f_0(500)}/\theta_{f_0(500)}^0=0.5-1.2$.
 The lowest masses of the $f_0(980)$ and $a_0(980)$ can reproduce the experimental values approximately;  while the lowest masses of the $\kappa_0(800)$ and $f_0(500)$ are larger than the   experimental values. In calculations, we observe that the minima of the  predicted masses vary with the Borel parameters $M^2$ and threshold parameters $s_S^0$, the mixing angles  $\theta_S^0$ are the best values.

In Fig.11, we plot the pole contributions  of the scalar mesons as the  two-quark-tetraquark mixed states  with variations of the  mixing  angles  $\theta_S$,
where the same parameters as  that in Fig.10 are taken. From the figure, we can see that the pole contributions increase with the $\theta_S/\theta_S^0$ slowly, and reach the   maxima  at the values $\theta_S/\theta_S^0=1.0-1.3$, then decrease quickly and reach zero approximately. The best  values appear at the vicinity of the $\theta_S^0$, not far way from the $\theta_S^0$.

We can draw the conclusion tentatively that the QCD sum rules favor  the ideal two-quark-tetraquark mixing angles $\theta^0_S$.

\begin{table}
\begin{center}
\begin{tabular}{|c|c|c|c|c|c|c|c|}\hline\hline
                & $M^2 (\rm{GeV}^2)$ & $s_0 (\rm{GeV}^2)$   & pole             \\ \hline
   $f_0(980)$   & $0.8-1.2$          & $1.5\pm0.1$          & $(25-52)\%$     \\ \hline
   $a_0(980)$   & $0.8-1.2$          & $1.8\pm0.1$          & $(39-69)\%$     \\ \hline
$\kappa_0(800)$ & $0.6-1.0$          & $1.0\pm0.1$          & $(20-51)\%$     \\ \hline
   $f_0(500)$   & $0.6-1.0$          & $1.0\pm0.1$          & $(24-59)\%$     \\ \hline
 \hline
\end{tabular}
\end{center}
\caption{ The Borel parameters (or Borel windows), continuum threshold parameters and pole contributions of the QCD sum rules for the scalar nonet mesons as the two-quark-tetraquark mixed states. }
\end{table}

\begin{table}
\begin{center}
\begin{tabular}{|c|c|c|c|c|c|c|c|}\hline\hline
                &    & $m_{S}(\rm{GeV})$         & $\lambda_{S}(10^{-4}\rm{GeV}^5)$ \\ \hline
   $f_0(980)$   &    & $0.98\pm0.06$             & $8.7\pm1.3$              \\ \hline
   $a_0(980)$   &    & $0.97\pm0.05$             & $5.0\pm1.7$              \\ \hline
$\kappa_0(800)$ &    & $0.80\pm0.05$             & $3.6\pm0.6$              \\ \hline
   $f_0(500)$   &    & $0.70\pm0.06$             & $5.8\pm1.0$              \\ \hline
 \hline
\end{tabular}
\end{center}
\caption{ The  masses and pole residues of the scalar nonet mesons as the  two-quark-tetraquark mixed states. }
\end{table}

Now we study the finite width effects on the predicted masses. For example, the   currents $J_{f_0/a_0(980)}(x)$   couple potentially  with the scattering states  $K\bar{K}$, we take into account  the contributions of the  intermediate   $K\bar{K}$-loops to the correlation functions $\Pi_{f_0/a_0(980)}(p^2)$,
\begin{eqnarray}
\Pi_{f_0/a_0(980)}(p^2) &=&-\frac{\widehat{\lambda}_{f_0/a_0(980)}^{2}}{ p^2-\widehat{m}_{f_0/a_0(980)}^2-\Sigma_{K\bar{K}}(p)}+\cdots \, ,
\end{eqnarray}
where the $\widehat{\lambda}_{f_0/a_0(980)}$ and $\widehat{m}_{f_0/a_0(980)}$ are bare quantities to absorb the divergences in the self-energies $\Sigma_{K\bar{K}}(p)$.
All the renormalized self-energies  contribute  a finite imaginary part to modify the dispersion relation,
\begin{eqnarray}
\Pi_{f_0/a_0(980)}(p^2) &=&-\frac{\lambda_{f_0/a_0(980)}^{2}}{ p^2-m_{f_0/a_0(980)}^2+i\sqrt{p^2}\Gamma(p^2)}+\cdots \, .
 \end{eqnarray}
The contributions of the  other intermediate   meson-loops to the correlation functions $\Pi_{S}(p^2)$ can be studied in the same way.

We can take into account the finite width effects by the following simple replacements of the hadronic spectral densities,
\begin{eqnarray}
\delta \left(s-m^2_{S} \right) &\to& \frac{1}{\pi}\frac{\sqrt{s}\,\Gamma_{S}(s)}{\left(s-m_{S}^2\right)^2+s\,\Gamma_{S}^2(s)}\, .
\end{eqnarray}

It is easy to obtain the masses,
\begin{eqnarray}
m_S^2 &=& \frac{\int_0^{s^0_S}ds \, s\,\frac{1}{\pi}\frac{\sqrt{s}\,\Gamma_{S}(s)}{\left(s-m_{S}^2\right)^2+s\,\Gamma_{S}^2(s)}\, \exp\left(-\frac{s}{M^2} \right)}{\int_0^{s^0_S}ds \,\frac{1}{\pi}\frac{\sqrt{s}\,\Gamma_{S}(s)}{\left(s-m_{S}^2\right)^2+s\,\Gamma_{S}^2(s)}\, \exp\left(-\frac{s}{M^2} \right)} \, ,
\end{eqnarray}
where
\begin{eqnarray}
\Gamma_{f_0(980)}(s)&=&\Gamma_{f_0(980)} \, , \nonumber\\
\Gamma_{a_0(980)}(s)&=&\Gamma_{a_0(980)} \, , \nonumber\\
\Gamma_{\kappa_0(800)}(s)&=&\Gamma_{\kappa_0(800)} \frac{m^2_{\kappa_0(800)}}{s} \, , \nonumber\\
\Gamma_{f_0(500)}(s)&=&\Gamma_{f_0(500)} \frac{m^2_{f_0(500)}}{s} \, ,
\end{eqnarray}
and the masses $m_S$ at the right side of Eq.(41) come from the QCD sum rules in Eq.(34), here we have added the factors $\frac{m^2_{\kappa_0(800)}}{s}$ and $\frac{m^2_{f_0(500)}}{s}$ considering the large widths of the $\kappa_0(800)$ and $f_0(500)$. The numerical results are shown explicitly in Fig.12. From Fig.12, we can see that the predicted masses $m_{f_0(980)}$ and $m_{a_0(980)}$ are modified slightly after taking into account the small widths   $\Gamma_{f_0(980)}$ and $\Gamma_{a_0(980)}$, the finite widths can be neglected safely; while the predicted masses $m_{\kappa_0(800)}$ and $m_{f_0(500)}$ are modified considerably with the largest mass-shifts $\delta m_{\kappa_0(800)}=-0.09\,\rm{GeV}$ and $\delta m_{f_0(500)}=-0.04\,\rm{GeV}$. Now the predicted masses from the QCD sum rules are
\begin{eqnarray}
 m_{\kappa_0(800)}&=&(0.71\pm0.05) \, \rm{GeV} \, , \nonumber\\
m_{f_0(500)}&=&(0.66\pm0.06) \, \rm{GeV} \, ,
\end{eqnarray}
which are much better than the values presented in Table 3 compared to the experimental data,
\begin{eqnarray}
 m_{\kappa_0(800)}&=&(682 \pm29 ) \, \rm{MeV} \, , \nonumber\\
m_{f_0(500)}&=& (400-550) \, \rm{MeV} \, ,
\end{eqnarray}
from the Particle Data Group \cite{PDG}.

\begin{figure}
 \centering
  \includegraphics[totalheight=5cm,width=7cm]{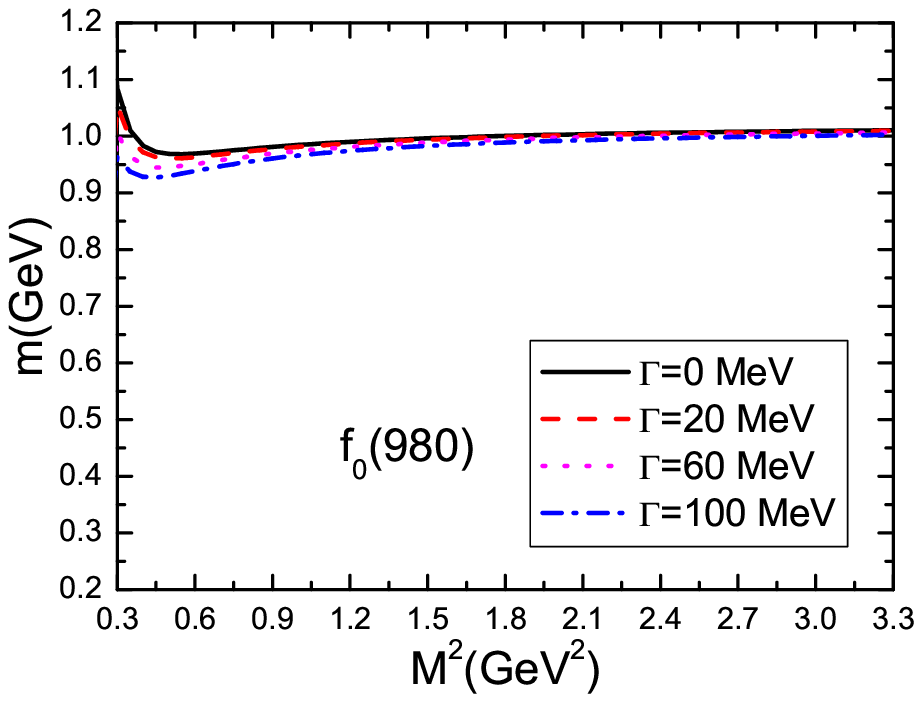}
 \includegraphics[totalheight=5cm,width=7cm]{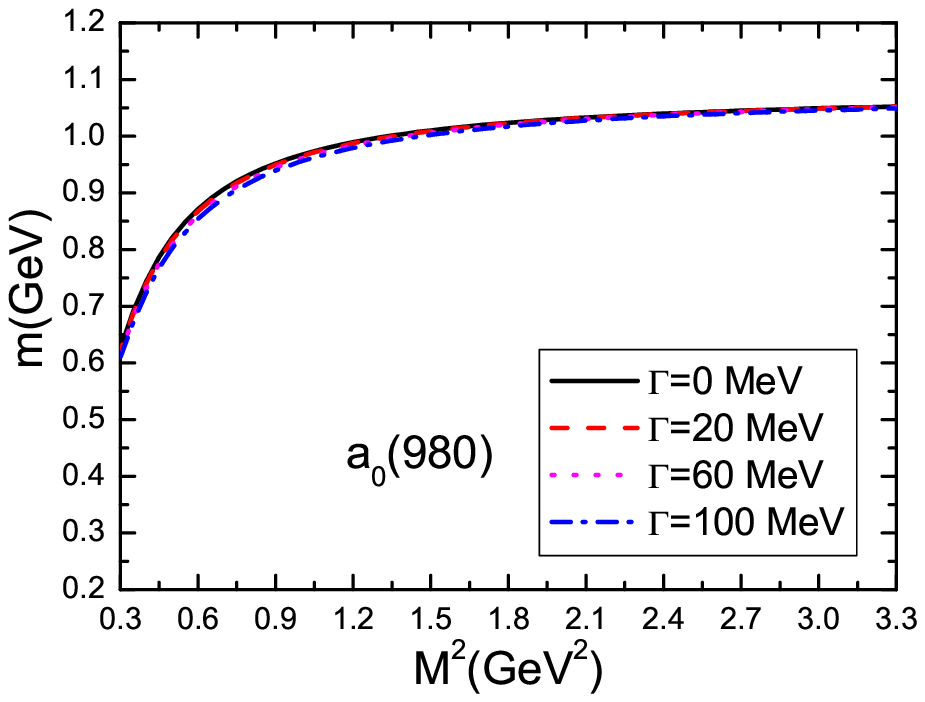}
 \includegraphics[totalheight=5cm,width=7cm]{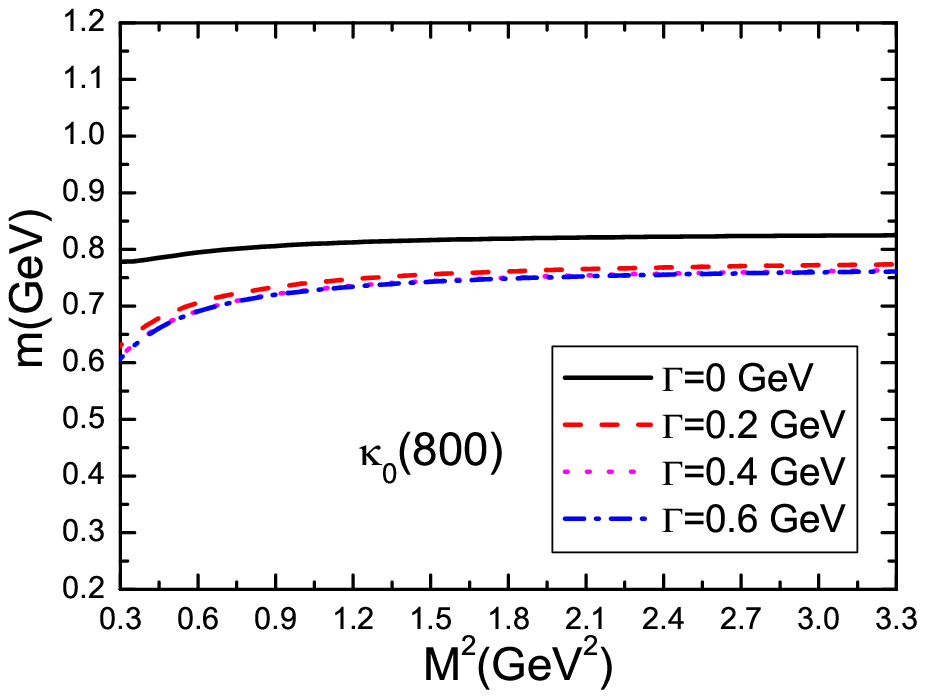}
 \includegraphics[totalheight=5cm,width=7cm]{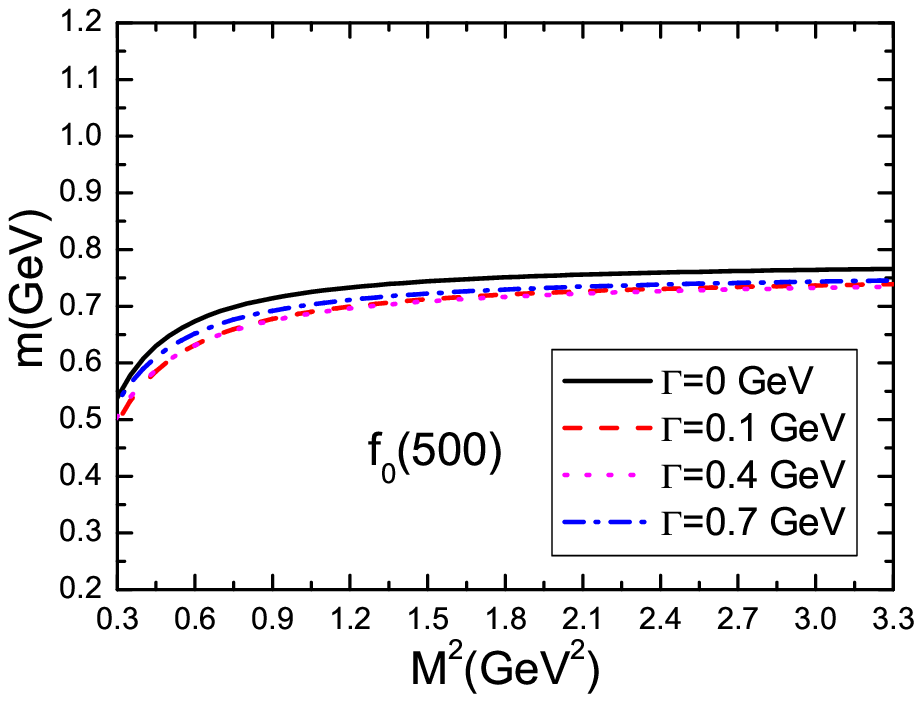}
    \caption{The masses of the scalar nonet mesons   with variations of the Borel parameters after taking into account the finite widths.}
\end{figure}

\section{Conclusion}
In this article, we assume that the nonet scalar mesons  below $1\,\rm{ GeV}$ are the two-quark-tetraquark mixed states and study their  masses and pole residues  using the  QCD sum rules. In calculation, we take into account the vacuum condensates up to dimension 10 and the $\mathcal{O}(\alpha_s)$ corrections to the perturbative terms, and neglect the condensates which are vacuum expectations   of the operators of the order $\mathcal{O}(\alpha_s^{>1})$,   in the operator product expansion.
We choose the ideal mixing angles,  which can lead to good convergent behavior in the operator product expansion,
the resulting  two-quark components are much larger than $50\%$. Then we  impose the two criteria (i.e. pole dominance and convergence of the operator product
expansion) of the QCD sum rules, search for the optimal  values of the Borel parameters  and continuum threshold
parameters, and obtain the masses and  pole residues of the nonet scalar mesons.
The predicted masses are compatible with  the experimental data, while the pole residues can be used to study the hadronic coupling constants and form-factors.

\section*{Acknowledgements}
This  work is supported by National Natural Science Foundation,
Grant Numbers 11375063, and Natural Science Foundation of Hebei province, Grant Number A2014502017.

\end{document}